%% file: sample-acmsmall-submission.tex
\documentclass[acmsmall,screen]{acmart}
\usepackage{cleveref} 
\usepackage{algorithm,algorithmicx}
\usepackage[noend]{algpseudocode}
\usepackage{subcaption}
\usepackage{multirow}

\begin{document}

\title{Plug-and-Hide: Provable and Adjustable Diffusion Generative Steganography}

\author{Jiahao Zhu}
\email{zhujh59@mail2.sysu.edu.com}
\affiliation{%
  \institution{School of Computer Science and Engineering, Sun Yat-sen University}
  \city{Guangzhou}
  \state{Guangdong}
  \country{China}
}
\author{Zixuan Chen}
\email{chenzx3@mail2.sysu.edu.cn}
\affiliation{%
  \institution{School of Computer Science and Engineering, Sun Yat-sen University}
  \city{Guangzhou}
  \state{Guangdong}
  \country{China}
}
\author{Jiali Liu}
\email{18879466196@163.com}
\affiliation{%
  \institution{Zhongshan School of Medicine, Sun Yat-sen University}
  \city{Guangzhou}
  \state{Guangdong}
  \country{China}
}
\author{Weiqi Luo}
\email{luoweiqi@mail.sysu.edu.cn}
\affiliation{%
  \institution{School of Computer Science and Engineering, Sun Yat-sen University}
  \city{Guangzhou}
  \state{Guangdong}
  \country{China}
}
\author{Yi Zhou}
\authornotemark[1]
\email{luoweiqi@mail.sysu.edu.cn}
\affiliation{%
  \institution{Zhongshan School of Medicine, Sun Yat-sen University}
  \city{Guangzhou}
  \state{Guangdong}
  \country{China}
}
\author{Xiaohua Xie}
\authornotemark[1]
\email{xiexiaoh6@mail.sysu.edu.cn}
\affiliation{%
  \institution{School of Computer Science and Engineering, Sun Yat-sen University}
  \city{Guangzhou}
  \state{Guangdong}
  \country{China}
}









\begin{abstract}
  Diffusion model-based generative image steganography (DM-GIS) is an emerging paradigm that leverages the generative power of diffusion models to conceal secret messages without requiring pre-existing cover images. In this paper, we identify a fundamental trade-off between stego image quality, steganographic security, and extraction reliability within the DM-GIS framework. Drawing on this insight, we propose \textbf{PA-B2G}, a \textbf{P}rovable and \textbf{A}djustable \textbf{B}it-to-\textbf{G}aussian mapping. Theoretically, PA-B2G guarantees the reversible encoding of arbitrary-length bit sequences into pure Gaussian noise; practically, it enables fine-grained control over the balance between image fidelity, security, and extraction accuracy. By integrating PA-B2G with probability-flow ordinary differential equations (PF-ODEs), we establish a theoretically invertible mapping between secret bitstreams and stego images. PA-B2G is model-agnostic and can be seamlessly integrated into mainstream diffusion models without additional training or fine-tuning, making it also suitable for diffusion model watermarking. Extensive experiments validate our theoretical analysis of the inherent DM-GIS trade-offs and demonstrate that our method flexibly supports arbitrary payloads while achieving competitive image quality and security. Furthermore, our method exhibits strong resilience to lossy processing in watermarking applications, highlighting its practical utility.
\end{abstract}

\maketitle

\input{Section/1_introcution}
\input{Section/2_relatedwork}

\input{Section/3_preliminary}
\input{Section/4_method}
\input{Section/5_exp}
\input{Section/6_conclusion}

\bibliographystyle{ACM-Reference-Format}
\bibliography{acmart}









\end{document}

%% file: Section/1_introcution.tex
\section{Introduction}
\label{sec:intro}
Steganography, as both an art and a science, aims to conceal secret data within seemingly innocuous carriers such as images for covert communication \cite{song2024survey,hu2024learning}. In traditional image steganography, benign images (\emph{a.k.a.} cover images) are usually selected as message carriers, and secret messages are embedded into their least significant bits using either non-adaptive embedding schemes \cite{luo2010edge} or adaptive embedding strategies \cite{filler2011minimizing}.
However, such algorithms generally suffer from limited message embedding capacity. Moreover, even at low payloads, message concealment inevitably introduces statistical discrepancies between cover images and their corresponding stego images, increasing the risk of being detected by machine learning-based steganalysis methods \cite{fridrich2012rich,xu2016structural,you2020siamese,wkk}.
  
To address these challenges, generative image steganography has emerged as a promising solution. Different from traditional image steganography, generative image steganography employs generative models to directly synthesize stego images from secret messages, thus avoiding the need for pre-defined cover carriers. Early approaches \cite{liu2017coverless,hu2018novel,wang2018sstegan,yu2021improved,liu2022image,wei2022generative,marobust,zhou2023generative,weiGSF,zhou2022secret} are mainly built upon Generative Adversarial Networks (GANs) \cite{goodfellow2020generative} and Flow models \cite{kingma2018glow}.
GAN-based methods \cite{liu2017coverless,hu2018novel,wang2018sstegan,yu2021improved,liu2022image,wei2022generative,marobust,zhou2023generative} usually require three collaborative networks: a generator, a discriminator, and a message extractor. This demands a delicate trade-off between image generation and message extraction during training, which raises training complexity and often results in unsatisfactory visual quality and message extraction accuracy.
In contrast, flow-based methods \cite{weiGSF,zhou2022secret} exploit the strict reversibility of flow models to establish a bijective mapping between secret messages and images. Nevertheless, this inherent reversibility constrains their performance in high-resolution image generation.

Recent studies \cite{yang2024gaussian,hu2024establishing,kim2025diffusion,zhou2025improved} leverage pre-trained image diffusion models \cite{rombach2022high,saharia2022photorealistictexttoimagediffusionmodels,ramesh2022hierarchical} for generative steganography. 
These methods typically follow a two-step pipeline: the first is an invertible bit-to-noise mapping that transforms secret bits into Gaussian or quasi-Gaussian noise; the second is stego image generation from the noise by solving a probability flow ordinary differential equation (PF-ODE) with solvers such as DPM-Solvers \cite{lu2022dpm,zheng2023dpm,lu2025dpm}. 
However, existing diffusion model-based generative image steganography (DM-GIS) methods struggle to flexibly support arbitrary-length secret payloads while simultaneously maintaining high message extraction accuracy, stego image quality, and steganographic security. 
Specifically, approaches in \cite{yang2024gaussian,hu2024establishing,zhou2025improved} achieve high extraction reliability and image fidelity but are limited to low effective payloads, which restricts their practicality. Diffusion-Stego \cite{kim2025diffusion} supports both low and high payloads but at the expense of compromised security and image quality. 
These limitations reveal two fundamental issues that remain underexplored: the intrinsic relationship among stego image quality, steganographic security, and message extraction accuracy, as well as the challenge of achieving an optimal trade-off among them. 
Although Kim \emph{et al.} empirically identifies a trade-off between image quality and extraction accuracy in DM-GIS \cite{kim2025diffusion}, it overlooks the critical dimension of steganographic security.

In this paper, we theoretically elucidate the underlying relationship among stego image quality, extraction accuracy, and steganographic security in DM-GIS, which is under-explored in previous studies. We identify that the key to balancing these three factors lies in the Gaussianity of the noise generated from secret bits. Specifically, DM-GIS methods that compromise Gaussianity are prone to producing sub-optimal stego images and achieving inferior security. Given a fixed encoding scheme and effective payload, pursuing higher message extraction accuracy inevitably impairs Gaussianity, thereby degrading generation quality and security. Based on these findings, we propose a provable and adjustable bit-to-Gaussian mapping, named PA-B2G, which reversibly transforms bit sequences of arbitrary length into pure Gaussian noise. PA-B2G consists of two stages: \textbf{i)} mapping bitstreams to uniformly sampled noise using a symmetric interval partitioning strategy; and \textbf{ii)} generating pure Gaussian noise via inverse transform sampling. Furthermore, we introduce non-sampling intervals and a variance-preserving adjustable algorithm, which enables fine-grained control over generation quality, security, and extraction accuracy. By iteratively denoising the noise generated by PA-B2G with an ODE solver, we establish a theoretically invertible mapping between secret messages and stego images. This approach allows PA-B2G to be integrated into most mainstream diffusion models without additional training or fine-tuning. Consequently, our method can be effectively extended to diffusion model watermarking.

The main contributions of this work are summarized as follows:
\begin{itemize}
\item We propose PA‑B2G, which can provably and reversibly transform arbitrary‑length bit sequences into standard Gaussian noise. Its adjustability enables a flexible and fine‑grained trade-off among stego image quality, steganographic security, and message extraction accuracy.
\item PA‑B2G decouples message embedding, stego image generation, and message extraction, which facilitates its direct integration into off-the-shelf diffusion models without any training or fine-tuning, thus realizing plug-and-hide functionality.
\item Experiments demonstrate that our method supports arbitrary-length payloads while achieving competitive stego image quality and steganographic security. Moreover, its resilience to lossy processing in watermarking highlights its practical utility.
\end{itemize}

The remainder of this paper is organized as follows.
\Cref{sec:relwork} reviews existing DM-GIS algorithms.
\Cref{sec:prelimilary} introduces preliminary knowledge of PF-ODE in diffusion models.
\Cref{sec:method} analyzes the relationship among stego image quality, steganographic security, and message extraction accuracy, and presents our provable and adjustable PA-B2G mapping. \Cref{sec:experiment} reports the experimental results. \Cref{sec:conclusion} concludes this work and discusses its limitations.

%% file: Section/2_relatedwork.tex
\section{Related Work}
GIS is a novel technique that conceals secret messages within synthesized images using generative models. It can be broadly categorized into three classes:
\label{sec:relwork}

\textbf{GAN-based Methods}. Liu \emph{et al.} \cite{liu2017coverless} pioneered GAN-based steganography by replacing class labels with secret data to synthesize stego images. Hu \emph{et al.} \cite{hu2018novel} employed GANs to convert secret data into noise vectors for generating carrier images. Wang \emph{et al.} \cite{wang2018sstegan} introduced a self-learning GAN-based approach for direct stego image generation from secret messages. Yu \emph{et al.} \cite{yu2021improved} proposed an attention-GAN method with optimized network architecture and loss functions.
Despite these improvements, such methods still suffer from issues including low image quality and unsatisfactory message extraction accuracy. To address these problems, Liu \emph{et al.} \cite{liu2022image} introduced an image disentanglement autoencoder for generative steganography, which improves both visual quality and extraction accuracy. Wei \emph{et al.} \cite{wei2022generative} incorporated a mutual information mechanism to enhance extraction accuracy and designed a hierarchical gradient decay strategy to strengthen anti-steganalysis performance.
To tackle robustness problems in message recovery and synthesis quality, Ma \emph{et al.} \cite{marobust} developed a weight modulation-based generator and a difference predictor to improve extraction robustness. Most recently, Zhou \emph{et al.} \cite{zhou2023generative} identified security vulnerabilities in the feature domain of existing GAN-based generative steganography schemes. They further proposed a framework that automatically generates semantic object contours to enhance steganographic security.

\textbf{Flow-based Methods}. To achieve higher embedding capacity and improved message extraction accuracy, Wei \emph{et al.} \cite{weiGSF} proposed a novel flow-based generative steganography algorithm, which directly embeds secret messages into the binarized initial noise of a pre-trained Glow model \cite{kingma2018glow}. In a similar vein, Zhou \emph{et al.} \cite{zhou2022secret} proposed Secret-to-Image Reversible Transformation (S2IRT), which incorporates the idea of permutation steganography permutation to encode secret bits by rearranging the order of entries in the latent vectors of Glow.

\textbf{Diffusion Model-based Methods}. With the advent of diffusion models, Peng \emph{et al.} \cite{peng2023stegaddpm} proposed StegaDDPM, a novel DM-GIS algorithm based on the denoising diffusion probabilistic model (DDPM) \cite{ddpm}. StegaDDPM embeds secret information by exploiting the probability distribution between the intermediate states and generated images of DDPM.
Yang \emph{et al.} \cite{yang2024gaussian} presented a provably secure bit-to-Gaussian mapping method using distribution-preserving sampling.
Hu \emph{et al.} \cite{hu2024establishing} introduced a novel bit-to-Gaussian mapping module based on orthogonal matrices and dual keys.
Zhou \emph{et al.} \cite{zhou2025improved} proposed a scheme named generative steganography
diffusion (GSD) that conceals secret bits in the frequency domain of the initial Gaussian noise.
Kim \emph{et al.} \cite{kim2025diffusion} designed four message projection strategies—Message to Noise (MN), Message to Binary (MB), Message to Centered Binary (MC), and Multi-bits—to balance image quality and message extraction performance.

%% file: Section/3_preliminary.tex
\section{Preliminary}
\label{sec:prelimilary}
Recent studies \cite{ddim, ddpm, karras2022elucidating, song2020score} have demonstrated that visually plausible generated images can be obtained by iteratively denoising  pure Gaussian noise. This denoising process can be modeled by the following PF-ODE \cite{song2020score, karras2022elucidating}:
\begin{equation}\label{eq:ode}
    \mathrm{d}\mathbf{x} = \left[ \mathbf{f}(\mathbf{x}, t) - \frac{1}{2} g(t)^2 \nabla_{\mathbf{x}} \log p_t(\mathbf{x}) \right] \mathrm{d}t,
\end{equation}
where $f(\mathbf{x},t)$ and $g(t)$ denote the drift term and diffusion coefficient with respect to timestep $t$, respectively. $\nabla_{\mathbf{x}} \log p_t(\mathbf{x})$ denotes the score function \cite{song2020score}, which points to the direction of higher probability density of the data $\mathbf{x}$ and is usually approximated by a deep neural network $\frac{1}{\sigma_t}\epsilon_\theta(\mathbf{x},t)$.
Assuming that the integration interval of \Cref{eq:ode} is set to $[0,T]$, an image $\mathbf{x}_0$ can be generated by integrating \Cref{eq:ode} from time $T$ to $0$; similarly, $\mathbf{x}_T$ can be recovered by integrating \Cref{eq:ode} along the reverse time direction. These two computational paths are fully invertible \cite{lai2025principles}. In practice, the integral associated with the denoising process can be approximately solved using off-the-shelf ODE solvers (e.g., the Euler method \cite{ddim} and DPM-Solvers \cite{lu2022dpm,zheng2023dpm}) by iteratively denoising pure Gaussian noise $\mathbf{x}_T\sim \mathcal{N}(\mathbf{0},\sigma_T^2\mathbf{I})$. This process can be expressed as:
\begin{equation}
    \mathbf{x}_{t-1} = \mathbf{x}_t + \int_t^{t-1} \left( f(t) \mathbf{x}_t + \frac{g(t)^2}{2\sigma_t} \boldsymbol{\epsilon}_\theta \left( \mathbf{x}_t, t \right) \right) \, \mathrm{d}t,
\end{equation}
Likewise, $\mathbf{x}_T$ can be approximately recovered by reversing this process. In this paper, we leverage high-order ODE solvers, such as the second-order Heun solver \cite{karras2022elucidating} and DPM‑Solver‑2 \cite{lu2022dpm}, since they introduce smaller local errors.

%% file: Section/4_method.tex
\section{Our Method}
\label{sec:method}
In \Cref{sec:exp-tradeoff}, we revisit the fundamental trade-off between stego image quality and message extraction accuracy, originally identified by Kim \emph{et al.} \cite{kim2025diffusion}, and further incorporate a discussion of steganographic security.
Building upon this analysis, we introduce the PA-B2G mapping in \Cref{sec:p-pa-b2g} and its adjustable variant in \Cref{sec:Enhancing}.
The detailed message hiding and extraction processes are then presented in \Cref{sec:gstego}.
\subsection{Theoretical Analysis}
\label{sec:exp-tradeoff}
\noindent \textbf{What Shapes Stego Image Quality and Security?} Let $\mathbf{g}_s$ and $\mathbf{g}$ denote the noise generated by DM-GIS and pure Gaussian noise, respectively, and let $\mathbf{x}_s$ and $\mathbf{x}$ denote the corresponding images generated by the diffusion model with parameter $\theta$. According to Theorem 1 proposed in \cite{zhu2026rethinking}, we have
\begin{equation}\label{eq:theorm1}
   D_{\text{KL}}(p(\mathbf{g}_s)\parallel p(\mathbf{g})) = D_{\text{KL}}(p_\theta(\mathbf{x}_s)\parallel p_\theta(\mathbf{x})).
\end{equation}
The optimization objective of diffusion models can be fundamentally formulated via maximum likelihood estimation \cite{ddpm} and KL divergence, as follows:
\begin{equation}\label{eq:mle}
    -E_{p_{\text{data}}} \left[ \log p_\theta(\mathbf{x}) \right] = D_{\text{KL}}\left( p_{\text{data}}(\mathbf{x}) \parallel p_\theta(\mathbf{x}) \right) + H\left( p_{\text{data}}(\mathbf{x}) \right),
\end{equation}
where $p_{\text{data}}$ and $p_\theta$ denote the real image distribution and the learned image distribution, respectively, and $H(\cdot)$ denotes the entropy function.
\Cref{eq:mle} implies that for a well-trained diffusion model, $D_{\text{KL}}\left( p_{\text{data}} \parallel p_\theta \right) = 0$.
Combined with \Cref{eq:theorm1}, we theoretically obtain
\begin{equation}\label{eq:theorm2}
   D_{\text{KL}}(p(\mathbf{g}_s)\parallel p(\mathbf{g})) = D_{\text{KL}}(p_\theta(\mathbf{x}_s)\parallel p_{\text{data}}(\mathbf{x})).
\end{equation}
\Cref{eq:theorm2} reveals that the Gaussianity of $\mathbf{g}_s$ directly affects the stego image quality as well as the steganographic security (see \cite{zhu2026rethinking} for details).
To achieve perfect stego image generation and security, one must ensure that $\mathbf{g}_s$ follows a pure Gaussian distribution.
Any steganographic operation that compromises the Gaussianity will theoretically degrade both visual quality and security. 

\noindent \textbf{What Shapes Message Extraction Accuracy?} As analyzed in \cite{zhu2026rethinking}, given a fixed encoding scheme and effective payload, one has to compromise the Gaussianity of $\mathbf{g}_s$, such as the adjustment schemes adopted in Diffusion-Stego \cite{kim2025diffusion} and S2IRT \cite{zhou2022secret}, to achieve higher message extraction accuracy. This inevitably compromises both generation quality and security, as demonstrated in \Cref{eq:theorm2}. Notably, to mitigate the destruction of Gaussianity, S2IRT employs an additional perturbation key in its message encoding function, which essentially breaks the constraint of ``under the same steganography encoding''.

\subsection{Design of PA-B2G}\label{sec:p-pa-b2g} As demonstrated in \Cref{sec:exp-tradeoff}, achieving perfect generation quality and security requires that the generated noise $\mathbf{g}_s$ be strictly Gaussian noise. To this end, we design a provably invertible bit-to-Gaussian mapping, termed PA-B2G, which comprises two stages. In the first stage, to handle secret bits with arbitrary lengths, the encrypted bit sequence $\mathbf{b}=b_1b_2\cdots b_n$ is divided into several subsequences of length $l$, forming an integer sequence $\mathbf{m}=m_1m_2\cdots m_k$ with $m_i\in\{0,1,\cdots,2^l-1\}$. Then, the integer sequence $\mathbf{m}$ is transformed into uniformly distributed noise $\mathbf{u}=u_1u_2\cdots u_k$, which can be defined by $\mathcal{C}_u(\mathbf{m})=\mathbf{u}$. Inspired by arithmetic coding, we leverage a symmetrical interval partitioning strategy and introduce the following two partition modes: 
\begin{itemize}
    \item[I.] The interval $[0,1]$ is evenly partitioned into $2^{l}$, and $u_i$ is sampled from $\mathcal{U}(\frac{m_i}{2^l}, \frac{(m_i+1)}{2^l})$.
    \item[II.] The interval $[0,1]$ is evenly partitioned into $2^{l+1}$, and $u_i$ is sampled from $\mathcal{U}(\frac{m_i}{2^{l+1}},\frac{m_i+1}{2^{l+1}})$ or $\mathcal{U}(1-\frac{m_i+1}{2^{l+1}}, 1-\frac{m_i}{2^{l+1}})$.
\end{itemize}

In the second stage, the obtained sequence $\mathbf{u}$ is transformed into pure Gaussian noise by the inverse transform sampling, which is defined by $\mathbf{g}_s=\mathcal{C}_g(\mathbf{u})$. $\mathcal{C}_g$ is also known as the percent point function (PPF) \emph{w.r.t.} $\mathcal{N}(0,1)$. The implementation details of PA-B2G and its inverse procedure are provided in \Cref{alg:pa-b2g} and \Cref{alg:pa-g2b}, respectively. Note that intervals partitioned by the two modes are all symmetric about $u=\frac{1}{2}$, which is the reason we refer to our method as the symmetrical interval partition, and its rationale is discussed in the next section. 

In practice, we can replicate and concatenate the secret message to generate the Gaussian noise $\mathbf{g}_s$. This approach essentially leverages repetition coding to improve extraction accuracy at the expense of effective embedding capacity. Consequently, as shown in \Cref{subsec:watermarking}, this trade-off is particularly suitable for diffusion model watermarking.

Here we first demonstrate the Gaussianity of the generated noise $\mathbf{g}_s$.
\begin{theorem}\label{proposition:p1}
    Given a message sequence $\mathbf{m}=m_1m_2\cdots m_k$,  $\mathbf{g}_s=\mathcal{C}_g\circ\mathcal{C}_u(\mathbf{m})\sim\mathcal{N}(\mathbf{0},\mathbf{I})$.
\end{theorem}
\begin{proof}
$p(g)\!\!=\!\!\sum\limits_{m}p(g|m)p(m)$. Since $\mathbf{m}$ is encrypted beforehand, $p(m)\!=\!2^{-l}$, and 
    \begin{equation}
        \label{proof:pro-e0}
        \begin{aligned}
            &p(g|m)=\int p(g,u|m)\mathrm{d}u=\int p(g|u,m)p(u|m)\mathrm{d}u\\
            &=p(g|\mathcal{C}_u(m))p(\mathcal{C}_u(m)).
        \end{aligned}
    \end{equation} 
    For the mode I, 
    \begin{equation}
        \label{proof:pro-e1}
        \begin{aligned}
            &p(g|m)=p(g|\mathcal{C}_u(m))p(\mathcal{C}_u(m))
            =\frac{2^lp(g|\mathcal{C}_u(m))}{(m\!+\!1\!-\!m)}\\
            & = \begin{cases} 
                \frac{2^l}{\sqrt{2\pi}} e^{-\frac{g^2}{2}} & \text{if } g \in [\mathcal{C}_g(\frac{m}{2^l}), \mathcal{C}_g(\frac{m+1}{2^l})] \\
                0 & otherwise.
            \end{cases}
        \end{aligned}
    \end{equation}
According to \Cref{proof:pro-e0},
\begin{equation}
    \begin{aligned}
        &p(g)=p(g|m)p(m)=\frac{1}{2^l}\sum\limits_m p(g|m)p(m)\\
        &=\sum\limits_m p(g|\mathcal{C}_u(m))=\frac{1}{\sqrt{2\pi}} e^{-\frac{g^2}{2}}:=\mathcal{N}(0,1), g\in \mathbb{R}.
    \end{aligned}
\end{equation}
For the mode II,
\begin{equation}
    \label{proof:pro-e2}
    \begin{aligned}
        &p(g|m)=p(g|\mathcal{C}_u(m))p(\mathcal{C}_u(m))=\frac{2^{l\!+\!1}p(g|\mathcal{C}_u(m))}{2(m\!+\!1\!-\!m)}\\
        & = \begin{cases} 
            \frac{2^l}{\sqrt{2\pi}} e^{-\frac{g^2}{2}} & \text{if } g \in [\mathcal{C}_g(\frac{m}{2^{l\!+\!1}}), \mathcal{C}_g(\frac{m\!+\!1}{2^{l\!+\!1}})], \\
            \frac{2^l}{\sqrt{2\pi}} e^{-\frac{g^2}{2}} & \text{if } g \in [\mathcal{C}_g(1-\frac{m\!+\!1}{2^{l+1}}), \mathcal{C}_g(1\!-\!\frac{m}{2^{l+1}})], \\
            0 & otherwise.
        \end{cases}
    \end{aligned}
\end{equation}
similarly, we can obtain $p(g)\!=\!\frac{1}{\sqrt{2\pi}} e^{-\frac{g^2}{2}}\!:=\!\mathcal{N}(0,1)$, where $g\in \mathbb{R}$. Since $g_i$ are \emph{i.i.d.}, $p(\mathbf{g})=\prod_{i=1}^{k}p(g_i)=\frac{1}{(2\pi)^{k/2}} e^{-\frac{1}{2}\mathbf{g}^T\mathbf{g}}:=\mathcal{N}(\mathbf{0}, \mathbf{I})$.
\end{proof}
\begin{algorithm}[!t]
    \caption{PA-B2G}
    \label{alg:pa-b2g}
    \begin{algorithmic}[1]
        \Require Secret message $\mathbf{m}$.
        \Ensure Pure Gaussian noise $\mathbf{g}$.
    \For {$i=1$ to $k$}
        \State Sample $r_i \sim \mathcal{U}(0,1)$;
            \State $u_i=\frac{r_i}
            {2^l}+\frac{m_i}{2^l}$; $\rhd$ mode I
            \State $u_i\!=\!\frac{r_i}{2^{l+1}}\!+\!\frac{m_i}{2^{l+1}}$ or $u_i=\frac{r_i}{2^{l+1}}+1-\frac{m_i+1}{2^{l+1}}$; $\rhd$ mode II
        \State $g_i=\mathcal{C}_g(u_i)$;
    \EndFor
    \State $\mathbf{g}=g_1g_2\cdots g_k$;
    \State \textbf{Return} $\mathbf{g}$.
    \end{algorithmic}
    \end{algorithm}
\begin{algorithm}[!t]
        \caption{Inverse Procedure of PA-B2G}
        \label{alg:pa-g2b}
        \begin{algorithmic}[1]
            \Require Noise $\mathbf{g}$.
            \Ensure Secret message $\mathbf{m}$.
        \For{$i=1$ to $k$}
        \State$u_i=\mathcal{C}^{-1}_g(g_i)$;
        \State$m_i=\lfloor u_i2^l\rfloor$; $\rhd$ mode I
        \State$m_i=\lfloor (1-u_i)2^{l+1}\rfloor$ \textbf{if} $u_i>0.5$ \textbf{else} $\lfloor u_i2^{l+1}\rfloor$; $\rhd$ mode II 
        \EndFor
        \State$\mathbf{m}=m_1m_2\cdots m_k$;
        \State\textbf{Return} $\mathbf{m}$.
        \end{algorithmic}
\end{algorithm}
\begin{figure}
  \centering
  \begin{subfigure}[b]{0.5\linewidth}
    \includegraphics[width=\linewidth]{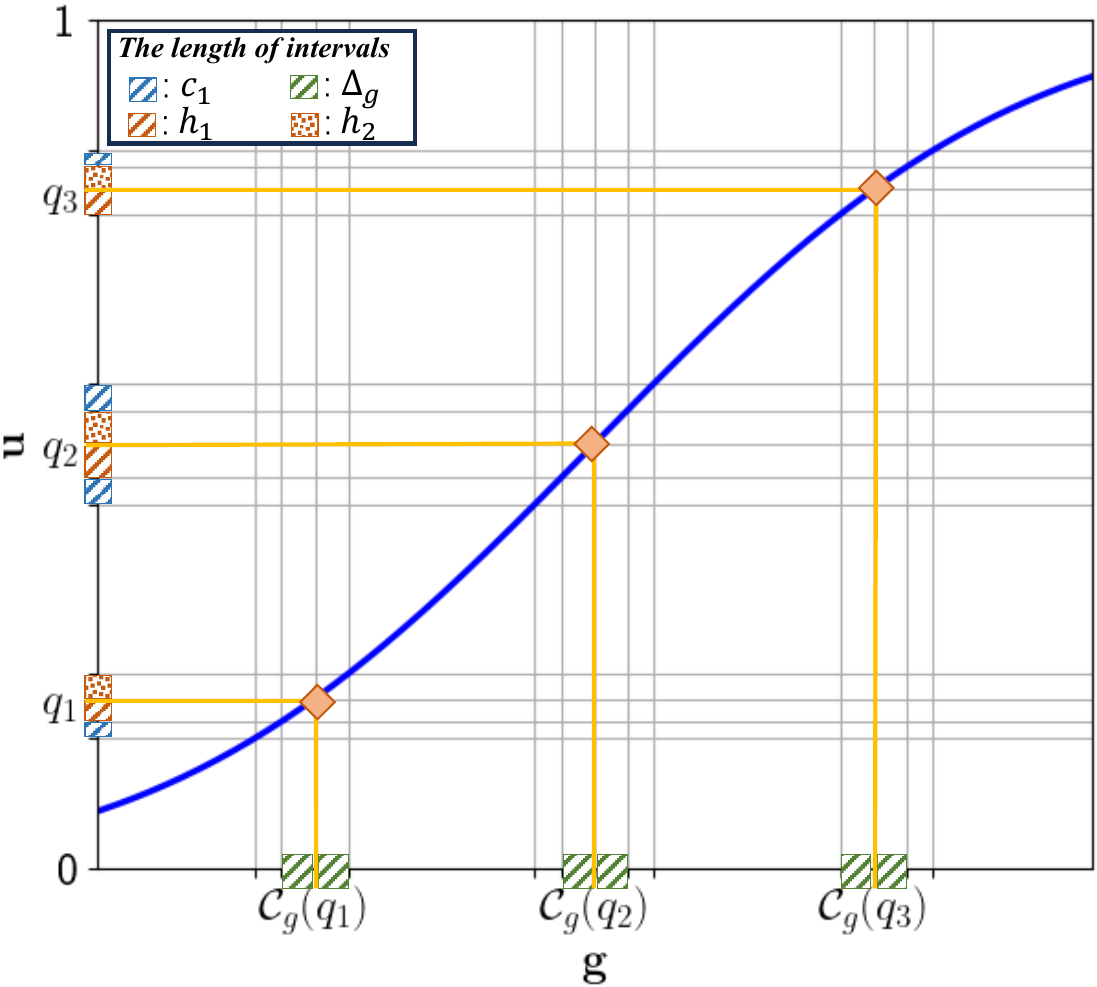}
    \caption{Variance increasing, mode I}
  \end{subfigure}\hfill
  \begin{subfigure}[b]{0.5\linewidth}
    \includegraphics[width=\linewidth]{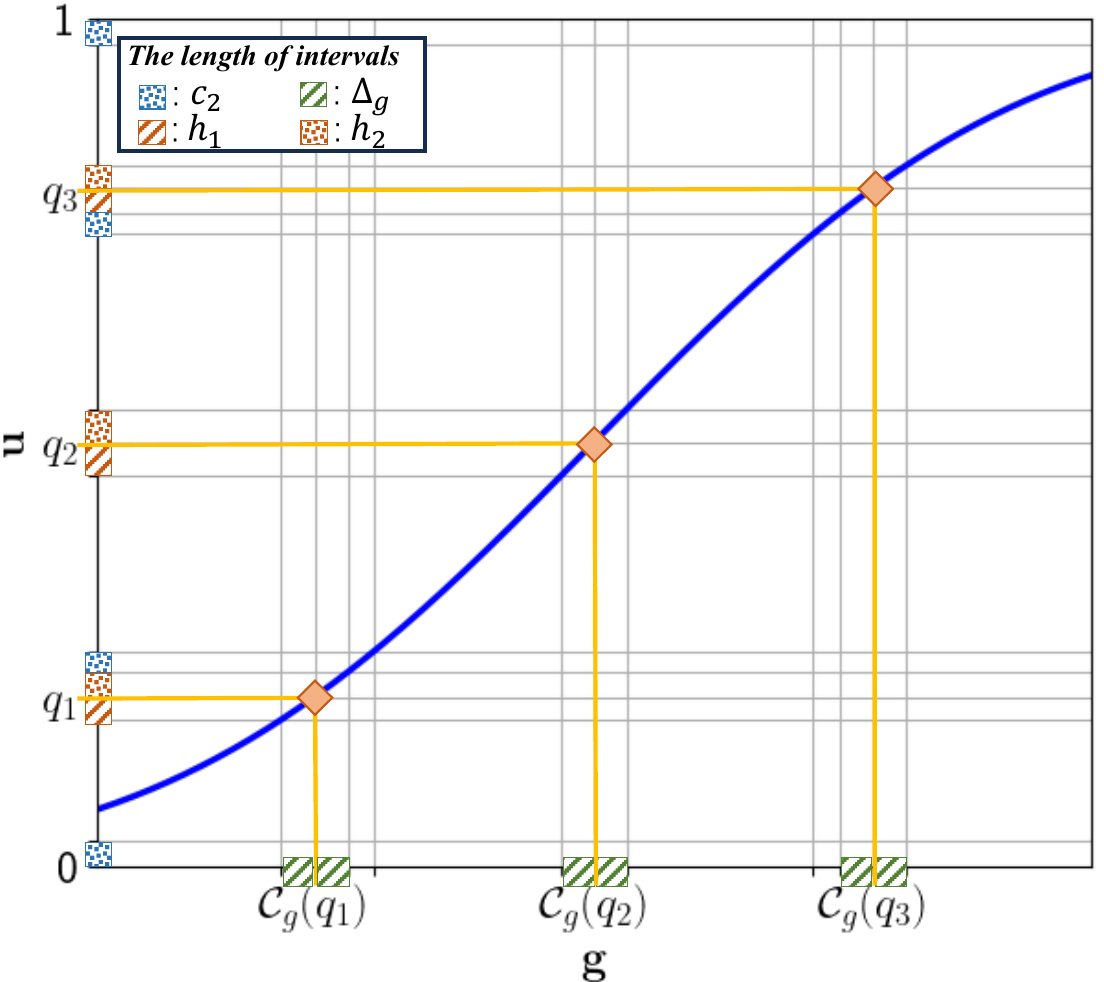}
    \caption{Variance decreasing, mode I}
  \end{subfigure}\hfill
  \begin{subfigure}[b]{0.5\linewidth}
    \includegraphics[width=\linewidth]{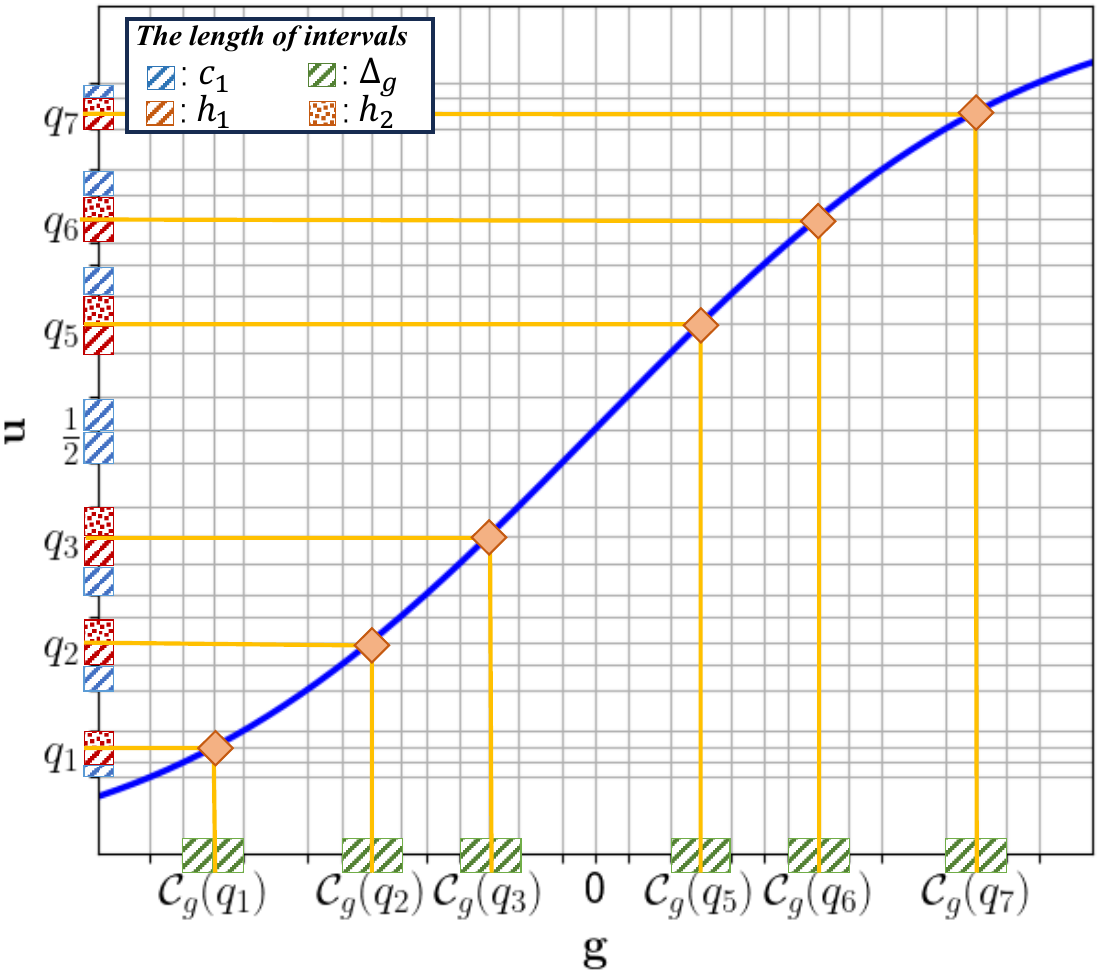}
    \caption{Variance increasing, mode II}
  \end{subfigure}\hfill
  \begin{subfigure}[b]{0.5\linewidth}
    \includegraphics[width=\linewidth]{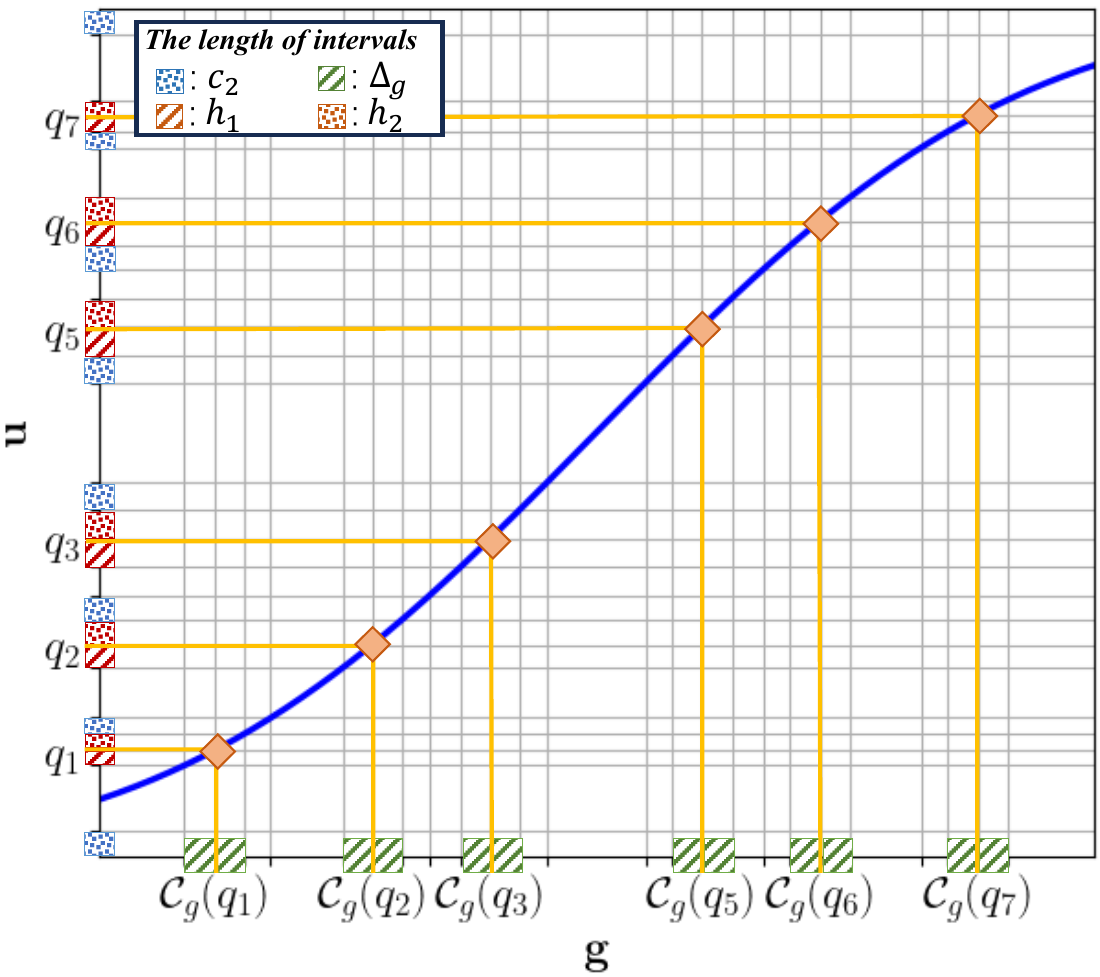}
    \caption{Variance decreasing, mode II}
  \end{subfigure}
  \caption{Examples of $l=2$ for modes I and II. Along the vertical axis in each figure, the red regions represent the neighborhoods of quantiles, and the blue regions represent the variance-correction intervals. Sampling within these regions is prohibited. The green regions are the intervals that we actually adjust. Our variance-preserving strategy in each iteration involves two key steps: (a) increasing $S^2_u$ by increasing $c_1$, and (b) decreasing $S^2_u$ by increasing $c_2$.}
    \label{fig:vp}
\end{figure}
\subsection{Adjustable PA-B2G}
\label{sec:Enhancing}
\Cref{sec:p-pa-b2g} demonstrates that PA-G2B can generate standard Gaussian noise and achieve provable security. However, in practical scenarios, numerical computation errors in ODE solvers and image quantization both significantly degrade message extraction accuracy. We observe that variations in $u_i$ within the $h$-neighborhoods of the quantiles (defined in \Cref{theorem:t2}) are more likely to induce extraction errors. Therefore, we choose to slightly compromise the Gaussianity of $\mathbf{g}_s$ to improve message extraction accuracy.

One straightforward yet effective solution is to avoid sampling within these neighborhoods. According to \cite{kim2025diffusion}, three conditions must be satisfied to achieve decent image generation performance: \textbf{i}) the sample mean of $\mathbf{g}_s$ must be close to $0$; \textbf{ii)} the sample variance  of $\mathbf{g}_s$ must be close to 1; \textbf{iii}) all entries of $\mathbf{g}_s$ must be independent and identically distributed. Motivated by these requirements, we propose a variance-preserving algorithm. Before presenting further details, we state the following preliminary results.
\begin{theorem}\label{theorem:t2}
	Let $
    \mathbf{q}=\{q_i|\frac{i}{2^l}, i=1,\cdots,2^l-1\}$ and $\tilde{\mathbf{q}}=\{\tilde{q}_i|\frac{i}{2^{l+1}}, i=1,\ldots,2^{l+1}-1, i\neq 2^l\}$ denote the quantile sets \emph{w.r.t.} partitioning mode I and II, respectively. Assuming that sampling in the $h$-neighborhoods of these quantiles, which are expressed as $(q_i-h,q_i+h)$ for mode I and $(\tilde{q}_i-h,\tilde{q}_i+h)$ for mode II, is prohibited, we have $\bar{g}=\frac{1}{k}\sum_{i=1}^{k}g_i=0$ and $S^2_g=\frac{1}{k-1}\sum_{i=1}^{k}(g_i-\bar{g})^2\neq 1$.
\end{theorem}
\begin{proof}
	According to \Cref{proposition:p1},
	\begin{equation}
        \label{theorem2:eq1}
		\begin{aligned}
		&p(g)=\sum\limits_mp(g|\mathcal{C}_u(m))p(\mathcal{C}_u(m))p(m) \\
		&=\frac{1}{2^l}\sum\limits_mp(g|\mathcal{C}_u(m))p(\mathcal{C}_u(m)).
		\end{aligned}
	\end{equation}
	For the mode I,
	\begin{equation}
        \label{theorem2:eq2}
            p(g)\!\!=\!\!\begin{cases}       
                0, &\!\!\!\!\!g\in\bigcup\limits_{i=1}^{2^l-1}(\mathcal{C}_g(q_i-h),\mathcal{C}_g(q_i+h)).\\
                \frac{2^l-h}{2^l\sqrt{2\pi}} e^{-\frac{g^2}{2}}, &\!\!\!\!\!g\!\!\in\!\!(\!-\infty,\mathcal{C}_g(q_1\!\!-\!\!h))\!\cup\!(\mathcal{C}_g(q_{2^l-\!1\!}+h)\!,\!+\infty),
                \\
                \frac{2^l-2h}{2^l\sqrt{2\pi}} e^{-\frac{g^2}{2}}, &\!\!\!\!\!otherwise.\\ 
		\end{cases}	
	\end{equation}
As for the mode II,
\begin{equation}
    \label{theorem2:eq3}
       p(g)\!\!=\!\!\begin{cases}       
            0, &\!\!\!\!\!g\in\bigcup\limits_{i=1,i\neq2^l}^{2^{l+1}-1}(\mathcal{C}_g(\tilde{q}_i-h),\mathcal{C}_g(\tilde{q}_i+h)).\\
            \frac{2^l-2h}{2^l\sqrt{2\pi}} e^{-\frac{g^2}{2}}, &\!\!\!\!\!\!
            \begin{aligned}
                &g\!\in\!(\!-\infty,\mathcal{C}_g(\tilde{q}_1\!\!-\!\!h))\!\cup\!(\mathcal{C}_g(\tilde{q}_{2^{l\!+\!1}\!-\!1}\!+\!h),\!+\infty\!)\\
                &\cup(\mathcal{C}_g(\tilde{q}_{2^l-1}\!+\!h),\mathcal{C}_g(\tilde{q}_{2^l+1}\!-\!h)),
            \end{aligned}
            \\
            \frac{2^l-4h}{2^l\sqrt{2\pi}} e^{-\frac{g^2}{2}}, &\!\!\!\!\!otherwise.\\ 
    \end{cases}	
\end{equation}
According to \Cref{theorem2:eq2} and \Cref{theorem2:eq3}, 
the following conclusions hold true for both modes I and II: \textbf{i)} $gp(g)$ is an odd function; \textbf{ii)} the integral interval is symmetric about the axis $g=0$; and \textbf{iii)} the functions $gp(g)$ is bounded and contains a finite number of jump discontinuities within the integral interval. Collecting these conditions, we can derive that $\mathrm{E}[g]=\int_{-\infty}^{\infty}gp(g)\mathrm{d}g=0$. According to the Law of Large Numbers, when $k\to\infty$, we have $E[g]=\bar{g}=0$. Moreover, $D[g]=E[g^2]-(E[g])^2=E[g^2]=\int_{-\infty}^{+\infty} g^2p(g)\mathrm{d}g$. Clearly, according to \Cref{theorem2:eq2} and \Cref{theorem2:eq3}, $D(g)=E[g^2]\neq1$, and therefore, when $k\to \infty$, $D(g)=S^2_g\neq 1$.
\end{proof}
\Cref{theorem:t2} indicates that the introduction of no-sampling intervals theoretically does not change the sample mean $\bar{g}$ of $\mathbf{g}$, primarily due to the symmetry of our interval partitioning strategy proposed in \Cref{sec:p-pa-b2g}. However, it induces a change in the sample variance $S_g^2$. Therefore, we only need to correct $S^2_g$ to satisfy the three conditions mentioned above. To this end, we design a variance-preserving algorithm, whose essence lies in iteratively adjusting the sampling intervals of $u$ to ensure that $S^2_g$ converges to 1. Specifically, if $S^2_g>1$, we iteratively reduce the sample variance $S^2_u$ of $\mathbf{u}$; otherwise, we iteratively increase $S^2_u$. \Cref{fig:vp} shows two examples for $l=2$ \emph{w.r.t.} modes I and II, respectively, illustrating how $S^2_u$ is increased or decreased during each iteration. The complete implementation of our adjustable PA-B2G is given in \Cref{alg:pa-b2g:adjust}, and its reverse process is identical to that in \Cref{alg:pa-g2b}. Combining \Cref{alg:pa-b2g:adjust} and \Cref{fig:vp}, we observe that MN, MB, MC, and Multi-bits \cite{kim2025diffusion} are in fact four special cases of our PA-B2G.
  \begin{algorithm}[!t]
    \caption{Variance-Preserving Algorithm.}
    \label{alg:pa-b2g:adjust}
    \begin{algorithmic}[1]
    \Require Secret message $\mathbf{m}$, maximum iteration steps $N_{max}$, and error tolerance $e$.
    \Ensure Sampling result $\mathbf{g}_s$.
    \State Initialize $c_1=c_2=0$;
    \State Generate a seed sequence $s_1\cdots s_k$.
    \For{$i=1$ to $k$}
    \State Sample $r_i\sim \mathcal{U}(0,1)$ with seed $s_i$;
    \State $\rhd$ mode I
    \State $h_1=\frac{m_i+1}{2^l}-\mathcal{C}^{-1}_g(\mathcal{C}_g(\frac{m_i+1}{2^l})-\Delta_g)$;
    \State $h_2=\mathcal{C}^{-1}_g(\mathcal{C}_g(\frac{m_i}{2^l})+\Delta_g)-\frac{m_i}{2^l}$;
    \State  $u_i=(\frac{1}{2^l}-h_1-h_2-c_1-c_2)r_i+\frac{m_i}{2^l}+h_2$;
    \State $u_i=u_i + c_1$ \textbf{if} $\frac{m_i}{2}<\frac{1}{2}$ \textbf{else} $u_i + c_2$;
    \State $\rhd$ mode II
    \State $h_1=\frac{(m_i+1) \ \mathrm{mod} \ 2^l}{2^{l+1}}-\mathcal{C}^{-1}_g(\mathcal{C}_g(\frac{(m_i+1) \ \mathrm{mod} \ 2^l}{2^{l+1}})-\Delta_g)$;
    \State $h_2=\mathcal{C}^{-1}_g(\mathcal{C}_g(\frac{m_i}{2^{l+1}})+\Delta_g)-\frac{m_i}{2^{l+1}}$;
    \State  $u_i=(\frac{1}{2^{l+1}}-h_1-h_2-c_1-c_2)r_i+\frac{m_i}{2^{l+1}}+h_2+c_1$ or $u_i=(\frac{1}{2^{l+1}}-h_1-h_2-c_1-c_2)r_i+1-\frac{m_i+1}{2^{l+1}}+h_2+c_2$;
    \EndFor
    \For{$step=1$ to $N_{max}$}
    \State $\mathbf{g}=\mathcal{C}_{g}(\mathbf{u})$ and calculate $S_g^2$;
    \If{1-$S_g^2>e$}
    \State $c_1 = c_1 + \Delta_c$;
    \ElsIf{$S^2_g-1>e$}
    \State $c_2 = c_2 + \Delta_c$;
    \Else{\ break;}
    \EndIf
    \State Repeat lines 3-13;
    \EndFor
    \State \textbf{Return} $\mathbf{g}_s$.
    \end{algorithmic}
    \end{algorithm} 
 
In \Cref{alg:pa-b2g:adjust}, we do not treat $h$ as a hyperparameter; instead, we utilize $\Delta_g$ to dynamically determine the neighborhoods (referred to as no-sampling intervals) around each quantile. The rationale behind this is as follows: If we model the quantile as a continuous variable $q$, first, as $q\to 0$ or $1$, $\mathcal{C}_g(q+h)-\mathcal{C}_g(q-h)$ gradually increases, which significantly disrupts the Gaussianity of $\mathbf{g}$. Second, when $g=\mathcal{C}_g(u)$ undergoes a small numerical perturbation $\Delta_g$, we observe that $\Delta_u \approx \frac{\mathbf{d}\mathcal{C}^{-1}_g}{\mathbf{d}g}\Delta_g$. Since $\mathcal{C}_g$ is the PPF of $\mathcal{N}(0,1)$, $\frac{\mathbf{d}\mathcal{C}^{-1}_g}{\mathbf{d}g}$ is a symmetric function that first increases and then decreases over its domain. Consequently, as $q\to 0$ or $1$, $\frac{\mathbf{d}\mathcal{C}^{-1}_g}{\mathbf{d}g}\to 0$, leading to $\Delta_u\to 0$. This implies that overly long no-sampling intervals are unnecessary for $q\to 0$ or $1$. Based on this insight, we directly introduce a hyperparameter $\Delta_g$ to adjust the deviation of $\mathbf{g}$ from $\mathcal{N}(\mathbf{0},\mathbf{I})$, thereby enabling adaptive control over the length of the no-sampling intervals. As shown in \cref{fig:vp} and \Cref{theorem:t2}, for a given $\Delta_g$, as $q\to 0$ or $1$, both $h_1$ and $h_2$ approach $0$, which aligns seamlessly with our theoretical perspective.
\subsection{Message Hiding and Extraction with PF-ODEs}
\label{sec:gstego}
After obtaining the noise $\mathbf{g}_s$ via PA-B2G, $\mathbf{g}_s$ can be reparameterized to accommodate selected diffusion models. The process of generating stego images from $\mathbf{g}_s$ essentially amounts to a numerical integration problem for ODEs. A variety of ODE solvers have been applied to diffusion models. In this paper, we primarily adopt the 2$^\mathrm{nd}$-order Heun solver \cite{karras2022elucidating} for stego image generation and initial noise retrieval for the following reasons: \emph{i)} \textbf{Universality}: \cite{karras2022elucidating} provides a generalized ODE formulation that is compatible with most existing diffusion models using different noise schedulers (e.g., DDIM \cite{ddim}, VP-SDE \cite{song2020score}, and VE-SDE \cite{song2020score}), without requiring additional training or fine-tuning. \emph{ii)} \textbf{Efficiency}: Higher-order ODE solvers typically reduce local errors but also incur increased computational costs. 2$^\mathrm{nd}$-order Heun solver can achieve a favorable computational trade-off.

However, in practice, directly adapting the 2$^\mathrm{nd}$-order Heun solver to diffusion models defined over discretized timesteps may lead to numerical calculation errors. For instance, when adopting the noise scheduler of DDPM, we have $\sigma(t) = t = \sqrt{1 - \bar{\alpha}_t}$, where $\bar{\alpha}_t$ denotes a transformed noise scheduler defined in \cite{ddpm}. This gives rise to a critical issue: when attempting to retrieve $\mathbf{g}$ from $\mathbf{x}_0$, the Heun sampler fails because the initial noise parameter $\sigma(0) = \sqrt{1 - \bar{\alpha}_0} = 0$ (see the deterministic sampler defined in \cite{karras2022elucidating} for details). To resolve this problem, we propose retrieving $\mathbf{g}_s$ from $\mathbf{x}_\epsilon$ instead of $\mathbf{x}_0$, where $\epsilon$ is a small constant—for example, $\epsilon = 10^{-6}$ as employed in this paper. This modification does not affect the generation of stego images but effectively enhances the stability of message extraction.

%% file: Section/5_exp.tex
\section{Experiments}
\label{sec:experiment}
\subsection{Experimental setup}
\textbf{Evaluation Metrics}. We assess PA-B2G from four aspects: visual quality, steganographic security, message extraction accuracy, and embedding capacity.
\begin{itemize}
    \item Visual quality is evaluated using the Fr\'echet Inception Distance (FID) score, computed across 50,000 stego images.
    \item Steganographic security is measured by detection accuracy ($Acc_s$) using the state-of-the-art steganalyzer UCNet \cite{wkk}, with 5,000 cover-stego image pairs for training and 1,000 pairs for testing. Note that the cover images are also generated by generative models but do not contain any hidden secret messages.
    \item Message extraction accuracy is measured by the average accuracy ($\overline{Acc}$), calculated over 1,000 stego images.
    \item Embedding capacity is measured in bits per pixel (bpp), calculated as $\frac{|\mathbf{b}|}{h\times w}$, where $h$ and $w$ represent the image height and width, respectively.
\end{itemize}

\textbf{Setup of PA-B2G}. For low payloads, we automatically pad secret bit sequences with either 0 or 1 to match their length to the size of the generated images. For the following experiments, we set the maximum iteration steps $N_{max}=100$ and the correction step size $\Delta_c=\frac{1}{3072}$.
\begin{figure*}
  \centering
  \begin{subfigure}[b]{0.32\linewidth}
    \includegraphics[width=\linewidth]{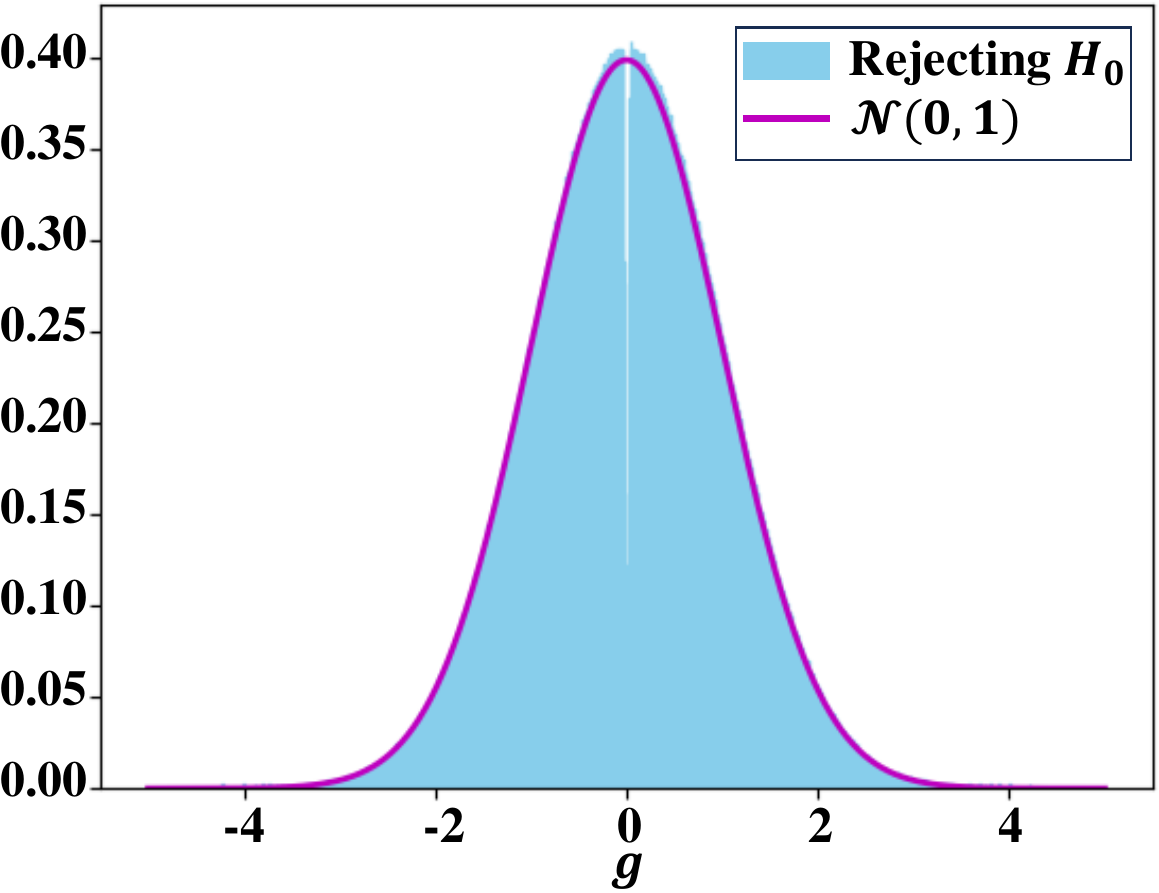}
    \caption{3 bpp, $128\times128$ mode I}
  \end{subfigure}\hfill
  \begin{subfigure}[b]{0.32\linewidth}
    \includegraphics[width=\linewidth]{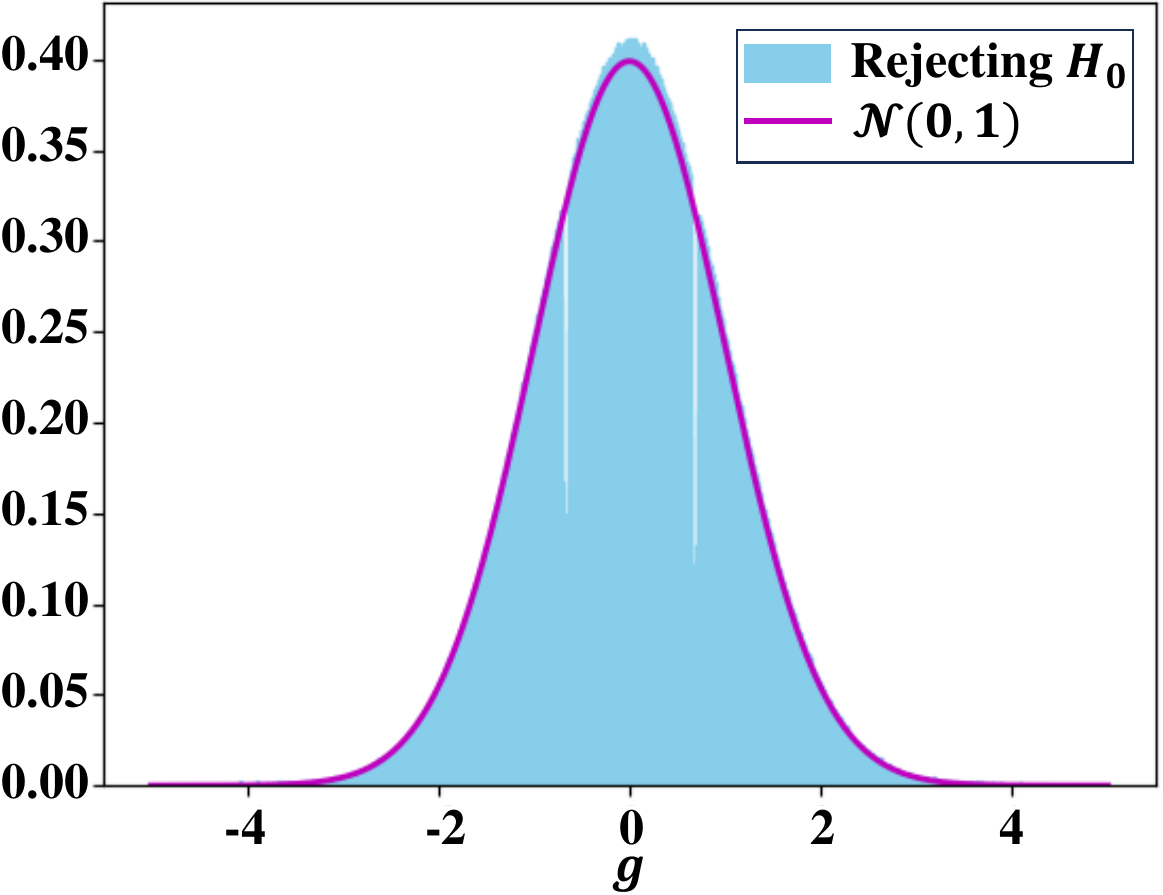}
    \caption{3 bpp, $128\times128$ mode II}
  \end{subfigure}\hfill
  \begin{subfigure}[b]{0.32\linewidth}
    \includegraphics[width=\linewidth]{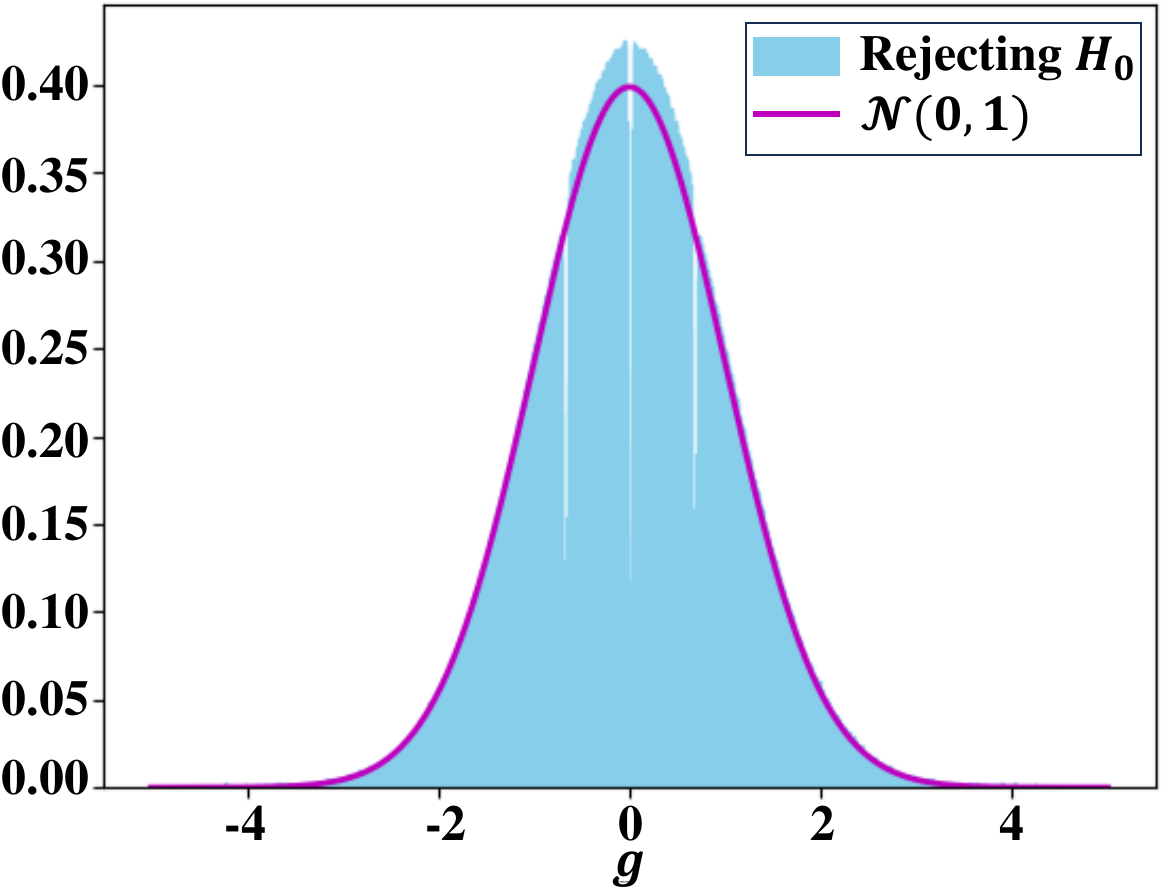}
    \caption{6 bpp, $128\times128$, mode I}
  \end{subfigure}
  \begin{subfigure}[b]{0.32\linewidth}
    \includegraphics[width=\linewidth]{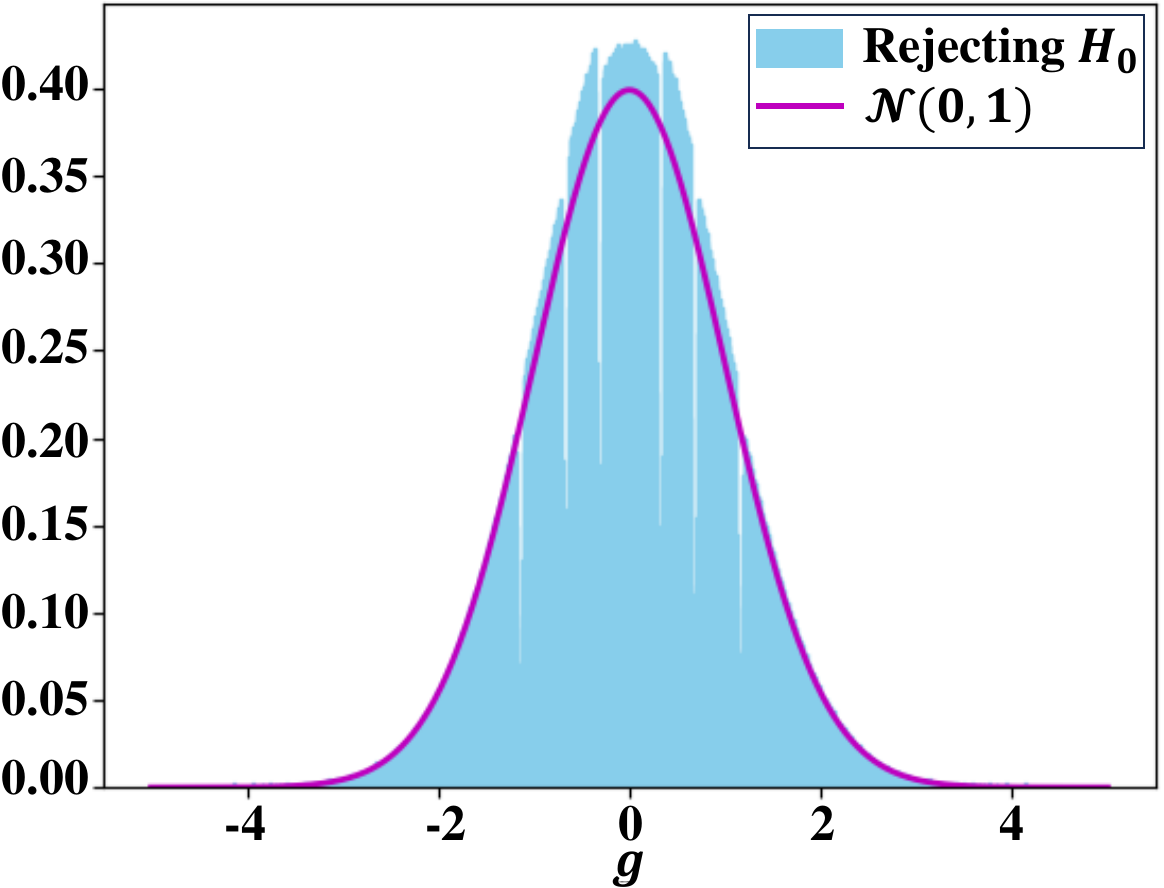}
    \caption{6 bpp, $128\times128$ mode II}
  \end{subfigure}\hfill
  \begin{subfigure}[b]{0.32\linewidth}
    \includegraphics[width=\linewidth]{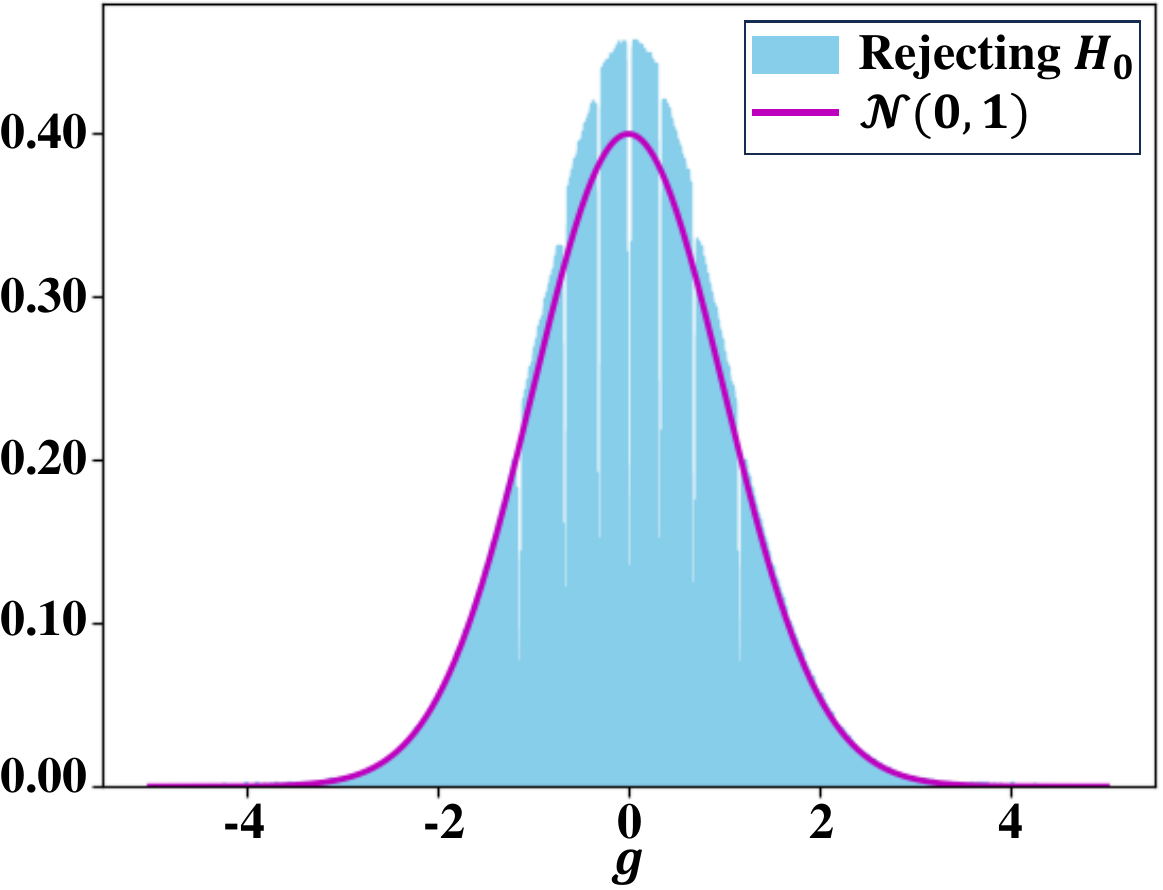}
    \caption{9 bpp, $128\times128$, mode I}
  \end{subfigure}\hfill
  \begin{subfigure}[b]{0.32\linewidth}
    \includegraphics[width=\linewidth]{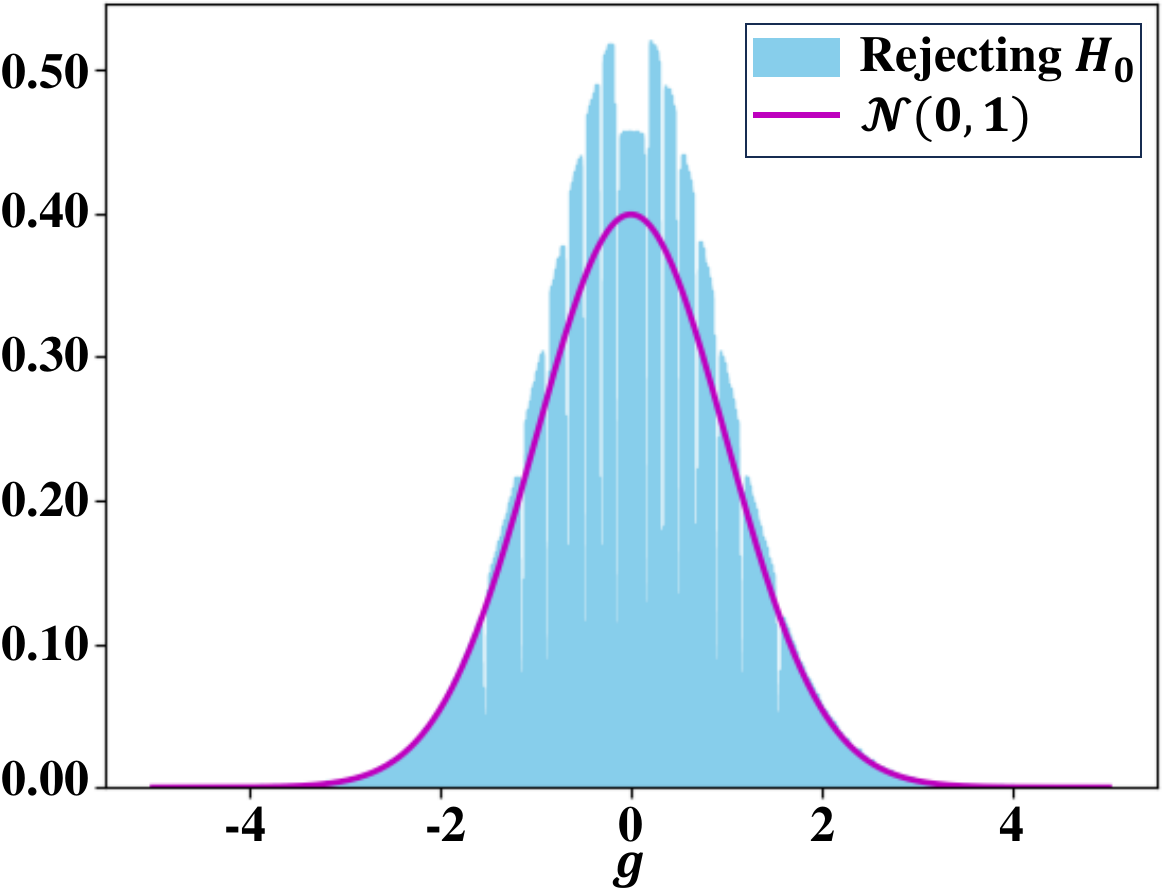}
    \caption{9 bpp, $128\times128$, mode II}
  \end{subfigure}
  \begin{subfigure}[b]{0.32\linewidth}
    \includegraphics[width=\linewidth]{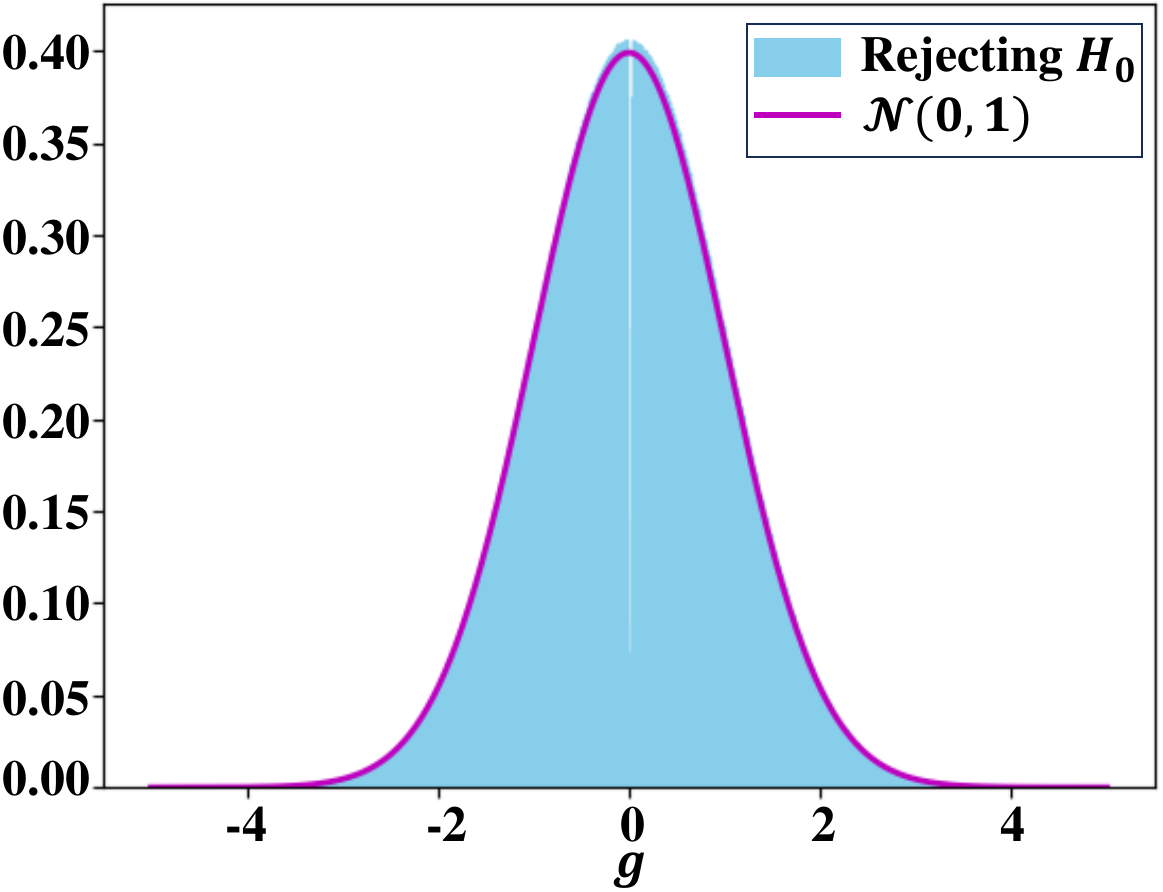}
    \caption{3 bpp, $256\times256$, mode I}
  \end{subfigure}\hfill
  \begin{subfigure}[b]{0.32\linewidth}
    \includegraphics[width=\linewidth]{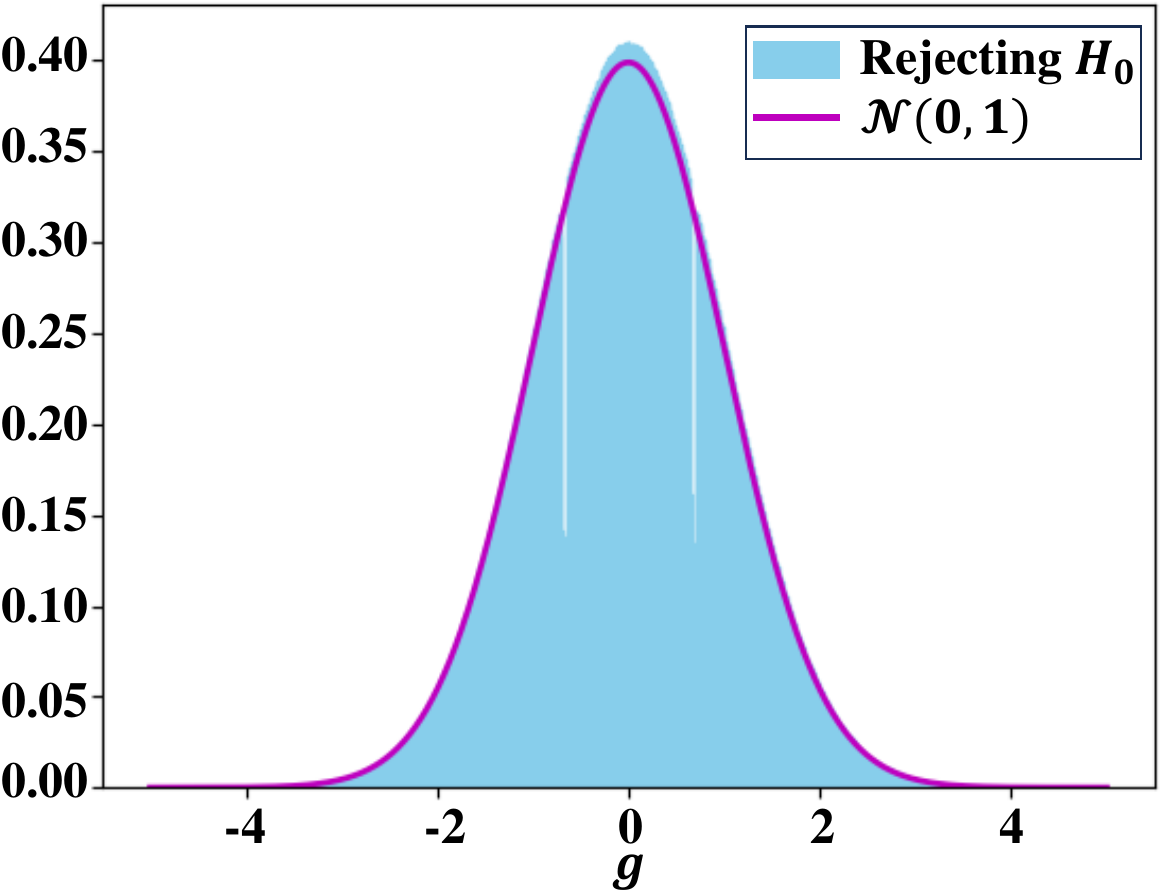}
    \caption{3 bpp, $256\times256$, mode II}
  \end{subfigure}\hfill
  \begin{subfigure}[b]{0.32\linewidth}
    \includegraphics[width=\linewidth]{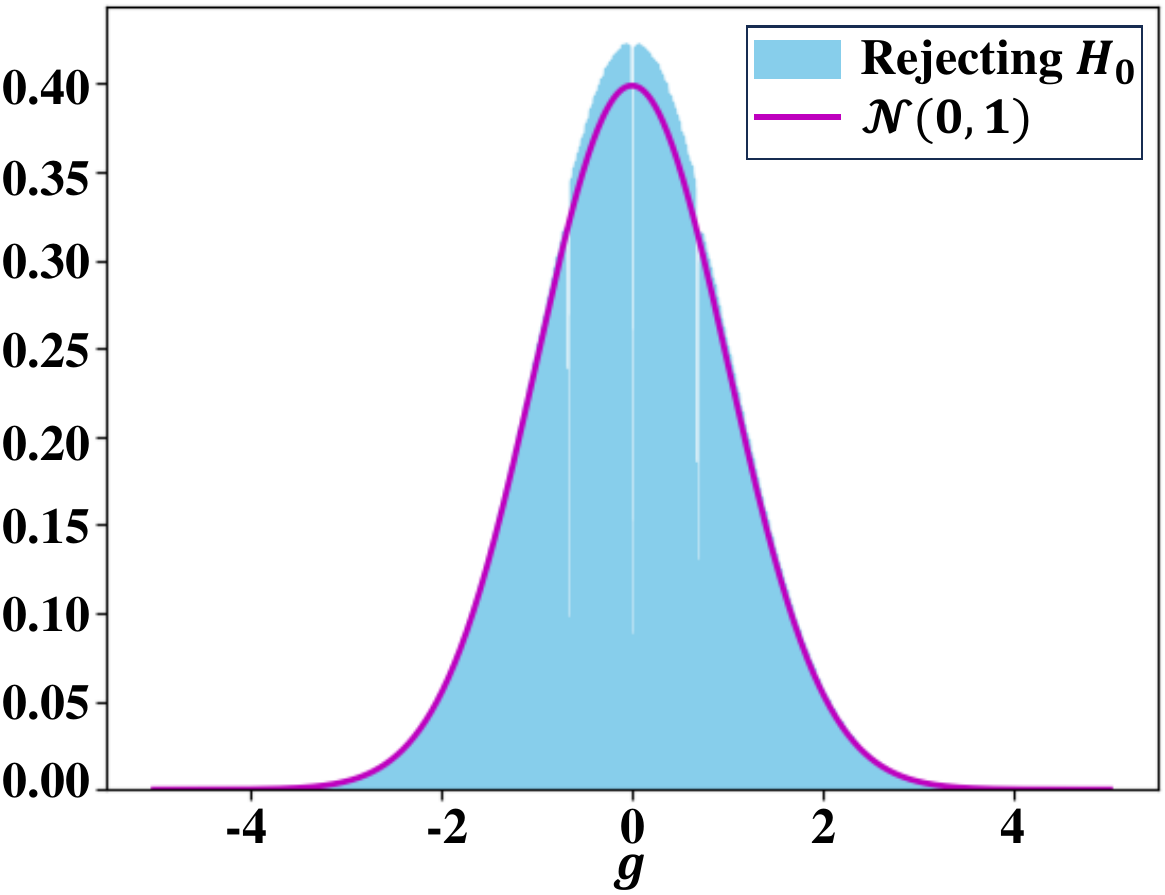}
    \caption{6 bpp, $256\times256$, mode I}
  \end{subfigure}
  \begin{subfigure}[b]{0.32\linewidth}
    \includegraphics[width=\linewidth]{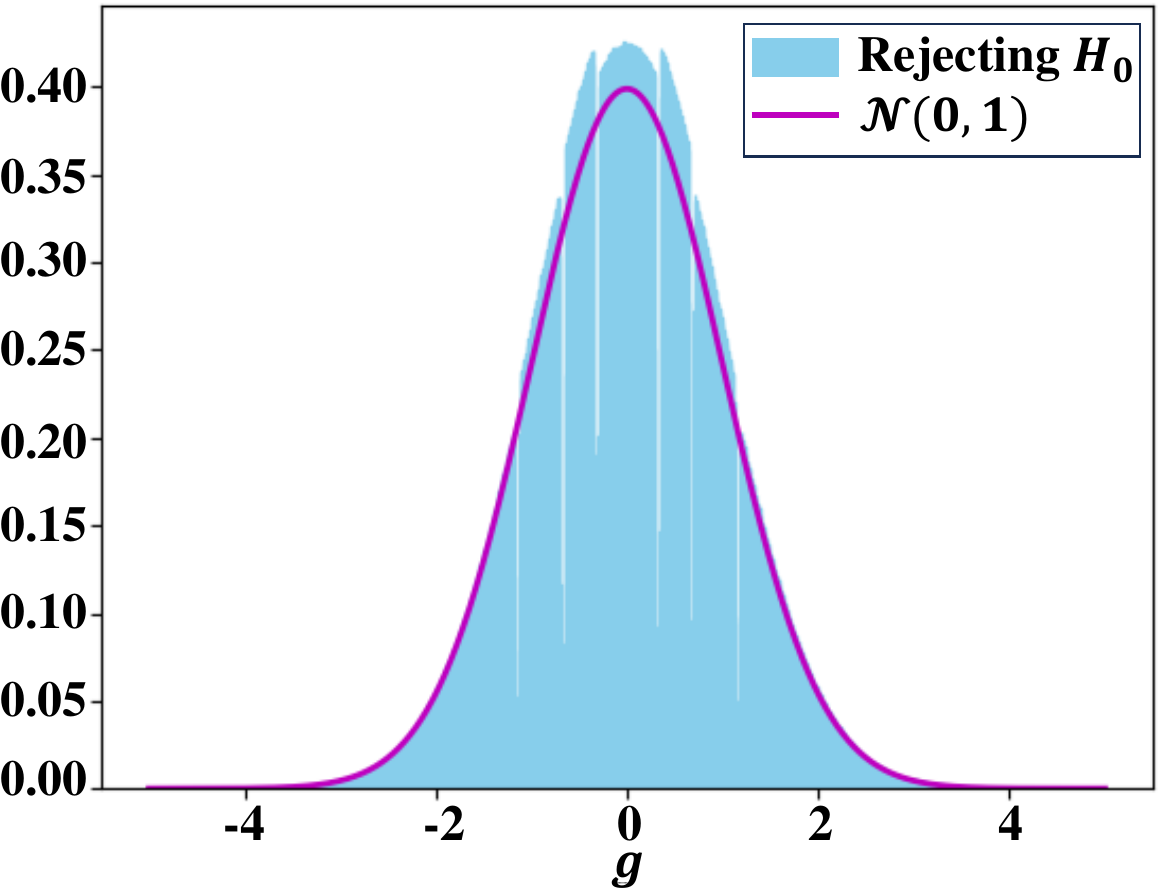}
    \caption{6 bpp, $256\times256$, mode II}
  \end{subfigure}\hfill
  \begin{subfigure}[b]{0.32\linewidth}
    \includegraphics[width=\linewidth]{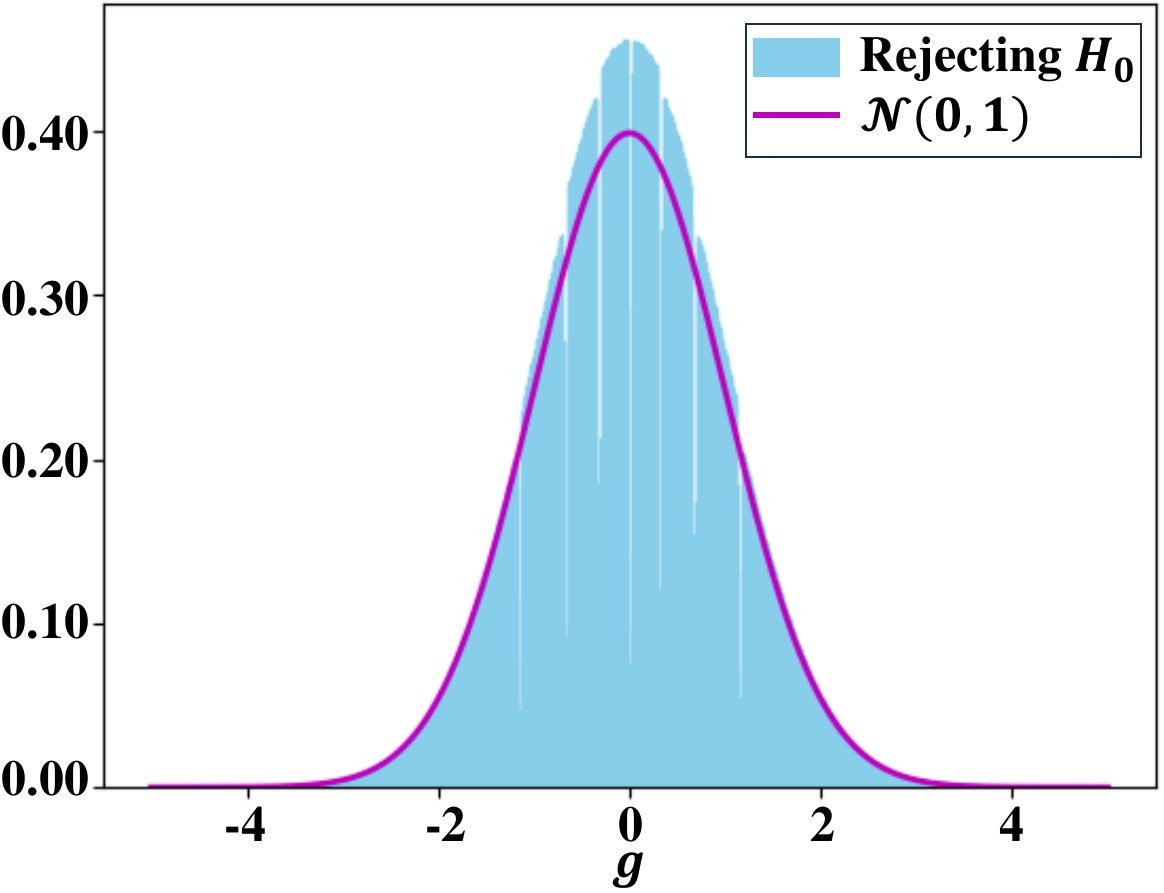}
    \caption{9 bpp, $256\times256$, mode I}
  \end{subfigure}\hfill
  \begin{subfigure}[b]{0.32\linewidth}
    \includegraphics[width=\linewidth]{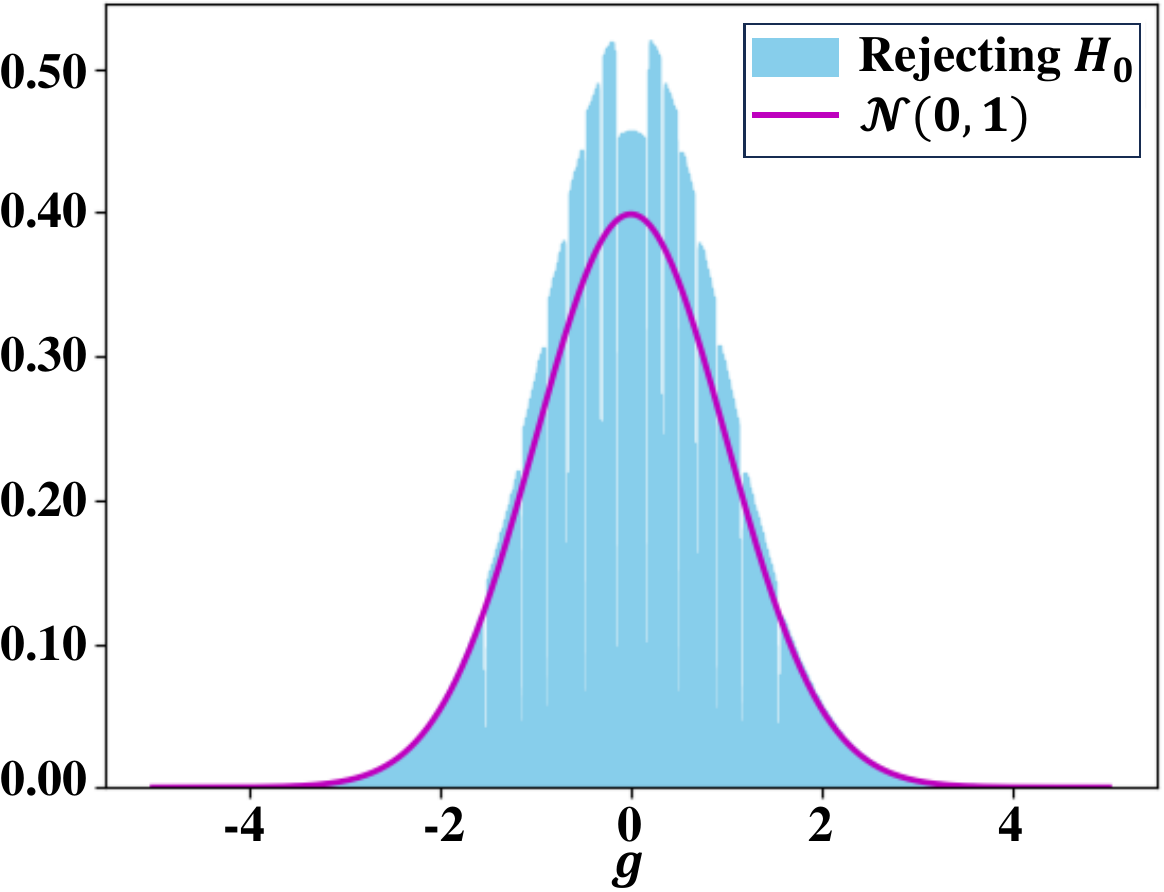}
    \caption{9 bpp, $256\times256$, mode II}
  \end{subfigure}

  \caption{Averaged histograms of noise that rejecting $H_0$ \emph{w.r.t.} different payloads and noise sizes under $\Delta_g=0.02$.}
  \label{fig:exp-nt}
\end{figure*}
 \begin{table*}[!t]
  \captionbox{Normality test of samples generated by PA-B2G in modes I and II with four different sample sizes \emph{w.r.t.} varying payloads and $\Delta_g$.}{
		\centering
				\begin{tabular}{clcc c cc c cc}
					\toprule
					&&\multicolumn{2}{c}{3 bpp}&& \multicolumn{2}{c}{6 bpp} && \multicolumn{2}{c}{9 bpp}\\
     \cmidrule{3-4}\cmidrule{6-7}\cmidrule{9-10}
                        \multicolumn{2}{c}{Noise Size($\times3$)}&Mode I & Mode II &&Mode I & Mode II &&Mode I & Mode II\\ \midrule
					\multirow{4}{*}{\rotatebox{90}{$\Delta_g\!=\!0.00$}}
					&$32\times 32$      &0.957 & 0.952&&0.959&0.954 && 0.952&0.951 \\ 
					&$64\times 64$      &0.949 & 0.947&&0.951&0.952 &&0.948&0.950 \\ 
					&$128\times 128$    &0.947 & 0.949 &&0.947&0.953 && 0.950&0.945 \\ 
					&$256\times 256$    &0.946 & 0.950 &&0.949&0.951 && 0.951&0.948 \\  
					\midrule
					\multirow{4}{*}{\rotatebox{90}{$\Delta_g\!=\!0.02$}}
					&$32\times 32$      &0.849 & 0.872&&0.835&0.844 && 0.845&0.795 \\ 
					&$64\times 64$      &0.405 & 0.601  &&0.431&0.401 && 0.393&0.333 \\ 
					&$128\times 128$    &0.00 & 0.00 &&0.00&0.00 && 0.00&0.00 \\ 
					&$256\times 256$    &0.00 & 0.00 &&0.00&0.00 && 0.00&0.00 \\ 
					\bottomrule
				\end{tabular}
    }
		\label{tab:normality}
	\end{table*}
\subsection{Normality Test}
We employ the one-sample Kolmogorov-Smirnov (K-S) test to assess the Gaussianity of samples generated by PA-B2G. For each sample, we test the null hypothesis $H_0$, which posits that the sample follows $\mathcal{N}(0,1)$, against the alternative hypothesis $H_1$. A $p$-value below $0.05$ leads to the rejection of $H_0$; otherwise, it is accepted. Through a random generation process, we create $1,000$ samples and compute the ratio $N_{H_0}/1,000$, where $N_{H_0}$ denotes the number of samples for which $H_0$ is accepted. \Cref{tab:normality} shows the test results \emph{w.r.t.} varying $\Delta_g \in \{0, 0.02\}$ and payloads of $3$~bpp, $6$~bpp, and $9$~bpp for both modes I and II. 

As observed, when no-sampling intervals are not introduced (\emph{i.e.}, $\Delta_g=0$), the samples generated by PA-B2G for any noise size and payload almost strictly adhere to $\mathcal{N}(0,1)$. However, even though $\Delta_g$ is a small value, the ratio  $N_{H_0}/1,000$ decreases significantly, particularly for large noise sizes. This may occur because the normality of samples generated by PA-B2G is highly sensitive to $\Delta_g$, while, conversely, the sensitivity of the K-S test improves as the sample scale increases. Furthermore, we depict the averaged histograms of noise samples that reject $H_0$ at varying payloads and noise sizes for $\Delta_g=0.02$ in \Cref{fig:exp-nt}. As illustrated, as the payload increases, the distribution of $\mathbf{g}_s$ gradually deviates from $\mathcal{N}(0,1)$. However, for low payloads, even when $H_0$ is rejected, these histograms still closely resemble the distribution of noise sampled from $\mathcal{N}(0,1)$.
 
\subsection{Method Comparison}
\begin{table*}[!t]\scriptsize
\captionbox{Comparison with existing message mapping methods using metrics FID, $Acc_s$, and, $\overline{Acc}$ on CIFAR-10 $32\times32$, FFHQ $64\times 64$, and LSUN-Bedroom $256\times256$ datasets under different payloads. PA-B2G$^1(.00)$ denotes PA-B2G in mode I with $\Delta_g=0.00$, and others follow similarly.}{
\setlength{\tabcolsep}{0.1mm}{
\begin{tabular}{lccccccccccccccccccccccc}
\toprule
  \multirow{4}{*}{Method} &  \multicolumn{7}{c}{CIFAR-10 $32\times32$} && \multicolumn{7}{c}{FFHQ $64\times64$} && \multicolumn{7}{c}{LSUN-Bedroom $256\times256$}\\\cmidrule{2-8}\cmidrule{10-16}\cmidrule{18-24}
                    &\multicolumn{3}{c}{3 bpp} && \multicolumn{3}{c}{6bpp}  &&  \multicolumn{3}{c}{3 bpp} && \multicolumn{3}{c}{6 bpp}  &&  \multicolumn{3}{c}{3 bpp} && \multicolumn{3}{c}{6 bpp}\\\cmidrule{2-4}\cmidrule{6-8}\cmidrule{10-12}\cmidrule{14-16}\cmidrule{18-20}\cmidrule{22-24}
                    &FID$\downarrow$ &$\overline{Acc}\uparrow$ &$Acc_s\downarrow$ &&FID$\downarrow$ &$\overline{Acc}\uparrow$ &$Acc_s\downarrow$&&FID$\downarrow$ &$\overline{Acc}\uparrow$  &$Acc_s\downarrow$&&FID$\downarrow$ &$\overline{Acc}\uparrow$ &$Acc_s\downarrow$ &&FID$\downarrow$ &$\overline{Acc}\uparrow$  &$Acc_s\downarrow$ &&FID$\downarrow$ &$\overline{Acc}\uparrow$ &$Acc_s\downarrow$\\ \midrule
MN \cite{kim2025diffusion}                  &2.02&83.86\%&50.60\%&&-&-&- &&2.46&86.22\%&50.83\%&&-&-&- &&  10.21         &   79.46\%           &  49.95\%        &&-            &-               &-\\
MB \cite{kim2025diffusion}                 &2.64            &94.36\%     & 92.35\%    &&-            &-     &-          &&4.16            &98.12\%      &99.70\%          &&-            &-     &-          && 10.98 &  \textbf{96.89\%}    &   99.90\%      &&-            &-     &-         \\
MC \cite{kim2025diffusion}                 &2.25            &84.96\%            & 67.00\% &&-            &-               &-          &&3.10            &92.70\%                &97.15\%          &&-            &-               &-          && 10.82           &  82.39\%    &   99.20\%       &&-            &-     &-         \\
Multi-bits \cite{kim2025diffusion}         &-            &-               &-  &&   2.37         &\textbf{83.01\%}             &    74.80\%      &&-            &-                &-          &&3.30            &\textbf{90.57\%}               &  98.85\%        &&-            &-      &-          &&    10.90        & \textbf{80.90\%}    &  99.90\%       \\
GSD \cite{zhou2025improved}      &2.98            &\textbf{98.89\%}     & 64.54\%&&-       &-     &-          &&3.08            &\textbf{98.81\%}                &50.24\%&&-            &-               &-         &&          10.89 & 94.11\%     &50.56\%&&-            &-     &-\\
GS \cite{yang2024gaussian}                  &1.99            &84.50\%               &50.94\%           &&2.01            &74.98\%               &50.12\% &&2.42            &86.24\%                &50.35\%          &&2.45            &79.98\%               &50.54\%          &&    10.21        &79.84\%      &50.74\% &&   10.22         &  72.04\%   &50.15\%         \\ \midrule
PA-B2G$^1$(.00) &\textbf{1.97}            &84.10\%               &\textbf{50.49\%}           &&2.03            &75.60\%               &50.75\% &&2.45            &86.85\%                &50.58\%          &&2.46            &79.25\%               &50.65\%          &&    \textbf{10.19}        &79.46\%      &49.74\% &&   \textbf{10.20}         &  71.12\%   &\textbf{49.50\%}         \\ 
PA-B2G$^1$(.08) &2.19            &88.16\%               &56.30\%           &&2.17            &80.02\%               &56.82\%          &&2.95            &92.76\%      &79.95\%          &&2.98            &88.40\%    &79.85\%          &&  10.75           &86.10\%      &    72.75\%      &&           10.78 & 78.10\%    & 72.35\% \\
PA-B2G$^2$(.00) &2.01            &75.64\%               &50.75\%           &&\textbf{2.00}            &69.11\%               &\textbf{50.43\%}          &&\textbf{2.40}            &79.22\%      &\textbf{50.43\%}          &&\textbf{2.43}            &70.90\%     &\textbf{50.55\%}          &&    10.21        &   72.14\%   &   \textbf{48.80\%}       &&     10.21       &   62.54\%   & 49.80\% \\
PA-B2G$^2$(.08) &2.13            &80.99\%               &  53.80\%         &&2.08            &71.53\%               &53.40\%          &&  2.94          &88.35\%      & 79.50\%         &&     2.89       &77.16\%     & 79.60\%          && 10.71          &  82.87\%    & 71.80\%         &&     10.73     & 71.90\%   & 72.30\%\\
\bottomrule
\end{tabular}}}
\label{tab:cmp_mm}
\end{table*}
\begin{table}[!t]
  \captionbox{Comparison with GAN-based and Flow-based GS methods using metrics FID, $Acc_s$, and, $\overline{Acc}$ on CelebA $64\times64$ under various payloads. PA-B2G$^1$(.08) denotes PA-B2G in mode I with $\Delta_g=0.08$.}{
		\centering
        \setlength{\tabcolsep}{0.4mm}
				\begin{tabular}{lccc c ccc c ccc}
					\toprule
					&\multicolumn{3}{c}{3 bpp}& &\multicolumn{3}{c}{6 bpp} && \multicolumn{3}{c}{9 bpp}\\
     \cmidrule{2-4}\cmidrule{6-8}\cmidrule{10-12}
                        Method&FID$\downarrow$&$\overline{Acc}\uparrow$&$Acc_s\downarrow$&&FID$\downarrow$&$\overline{Acc}\uparrow$&$Acc_s\downarrow$&&FID$\downarrow$&$\overline{Acc}\uparrow$&$Acc_s\downarrow$\\ \midrule
					GSF\cite{weiGSF}     &25.02 & \textbf{91.98\%}& 51.87\%&& 44.17 & 73.47\% &99.73\% && 51.80 & 62.85\%& 99.87\% \\ 
					GSN\cite{wei2022generative}     &16.11 & 74.31\%& 99.80\% &&17.02 & 60.75\%& 99.47\%&&19.98 & 52.25\%& 99.54\%  \\
                        S2IRT\cite{zhou2022secret}    & 44.52 &50.10\%&\textbf{50.85}\%&&45.74 &50.47\% &\textbf{50.12}\%&&- &-&- \\ 
					PA-B2G$^1$(.08)    &\textbf{7.70} & 90.81\%&76.75\% &&  \textbf{7.68}& \textbf{80.89\%}&76.54\% && \textbf{7.72}& \textbf{72.88\%}&\textbf{76.12}\%\\
					\bottomrule
				\end{tabular}}
	\label{tab:cmp_gf}
	\end{table}
\noindent\textbf{Comparison with Prior DM-GIS Methods}.
We compare PA-B2G with message mapping methods including MN, MB, MC, and Multi-bits proposed in~\cite{kim2025diffusion}, GSD~\cite{wei2023GSD}, and Gaussian Shading (GS) \cite{yang2024gaussian}. All message mapping methods are evaluated on CIFAR-10 $32\times32$, FFHQ $64\times64$, and LSUN-Bedroom $256\times256$ datasets. For each dataset, we equip each mapping method with the same generative backbone and sampling scheduler. Specifically, we employ the $2^\text{nd}$ order Heun ODE solver with 20 sampling steps on CIFAR-10 $32\times32$ and with 42 sampling steps on FFHQ $64\times64$, while using the $1^\text{st}$ order ODE solver proposed in DDIM \cite{ddim} with 18 sampling steps on LSUN-Bedroom $256\times256$. We set the error tolerance $e=0.0185$ defined in~\Cref{alg:pa-b2g:adjust} for CIFAR-10 $32\times32$ and FFHQ $64\times64$, and $e=0.007$ for LSUN-Bedroom $256\times256$. Before message hiding, we pad all secret bits and then encrypt them to meet required payloads. All mapping methods utilize the same random seed for stego image generation, and the corresponding generated images are saved in PNG format. Note that the payloads listed in \Cref{tab:cmp_mm} are effective payloads.
\begin{figure}
  \centering
  \begin{subfigure}[b]{0.33\linewidth}
    \includegraphics[width=\linewidth]{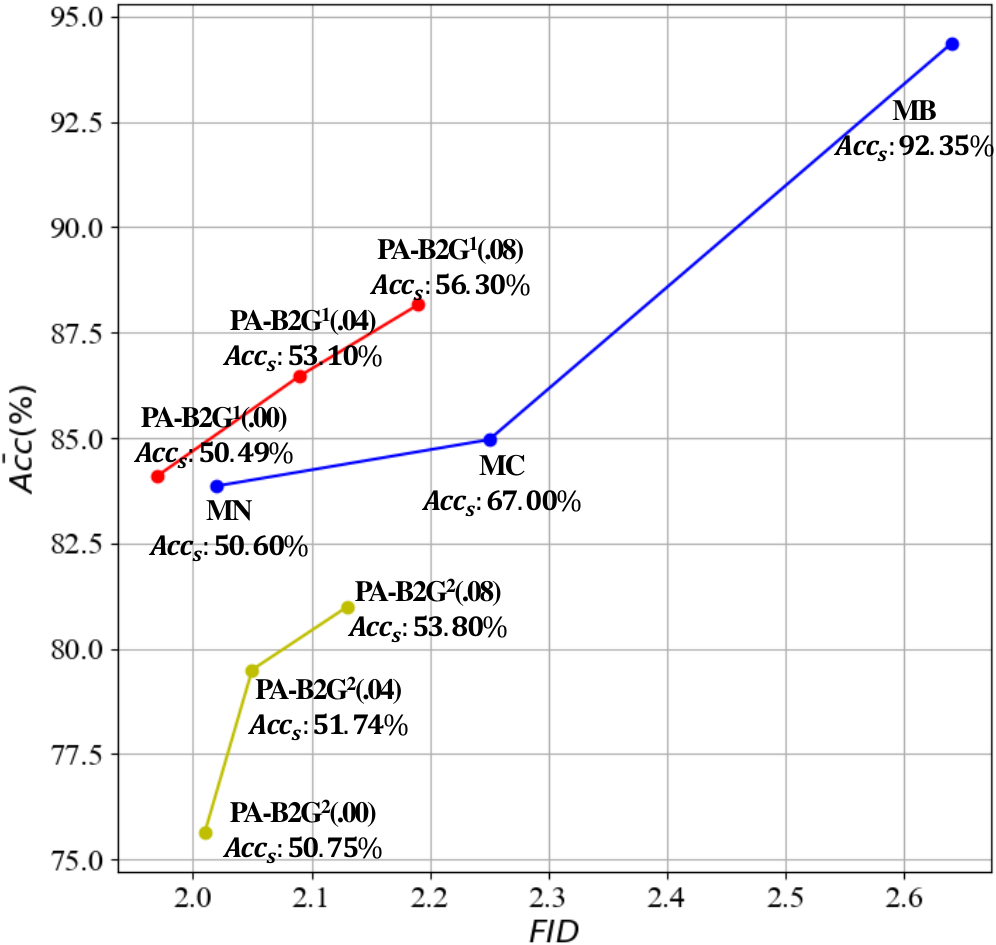}
    \caption{3 bpp, CIFAR-10 $32\times32$}
  \end{subfigure}\hfill
  \begin{subfigure}[b]{0.33\linewidth}
    \includegraphics[width=\linewidth]{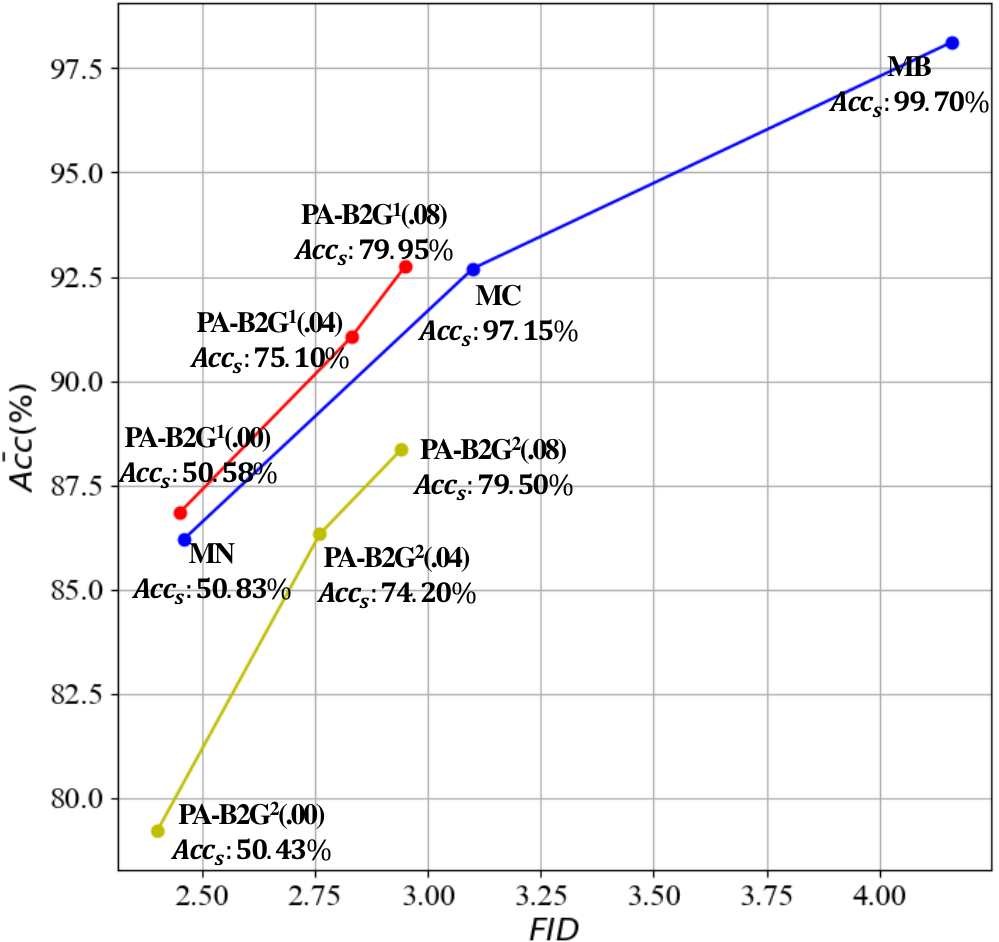}
    \caption{3 bpp, FFHQ $64\times64$}
  \end{subfigure}\hfill
  \begin{subfigure}[b]{0.32\linewidth}    \includegraphics[width=\linewidth]{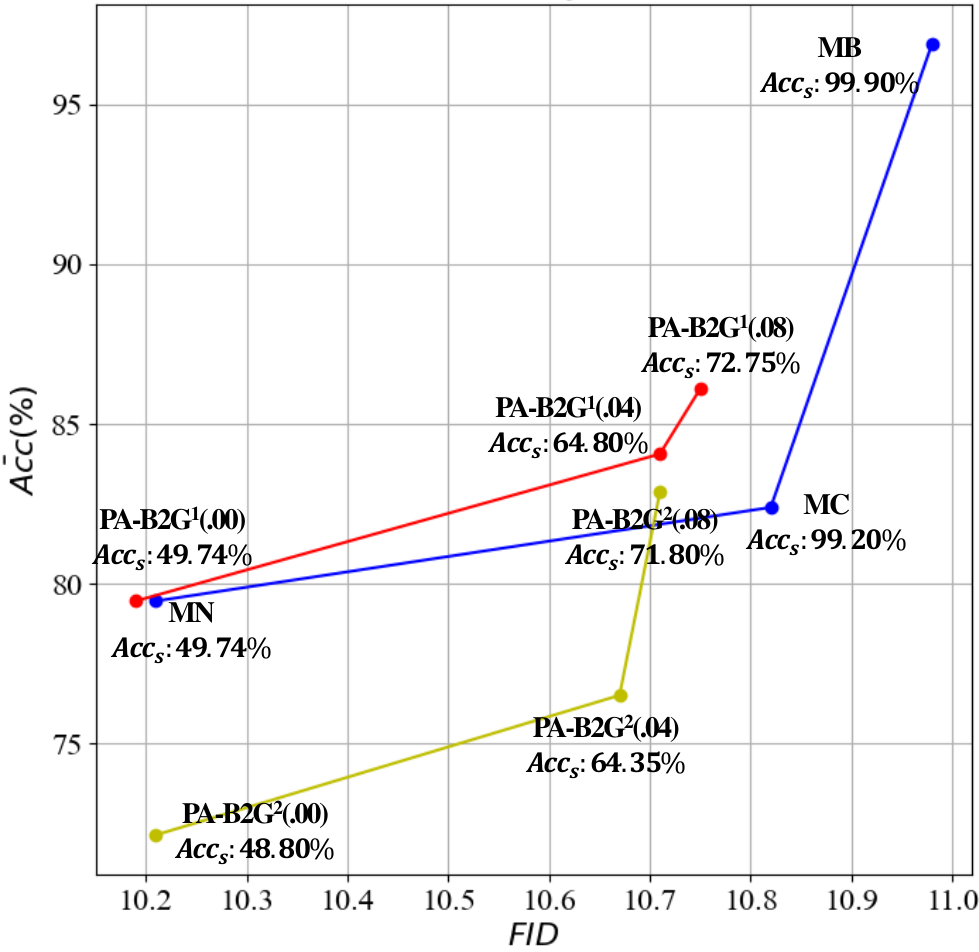}
    \caption{3 bpp, LSUN-Bedroom $256\times256$}
  \end{subfigure}\hfill
  \caption{Visualization results of the adjustment strategies adopted by \cite{kim2025diffusion} and PA-B2G on CIFAR-10, FFHQ, and LSUN-Bedroom.}
    \label{fig:trade-off}
\end{figure}

We list the FID score, average message extraction accuracy $\overline{Acc}$, and detection accuracy $Acc_s$ for each method across various payloads in \Cref{tab:cmp_mm}. MN, GS, and PA-B2G with $\Delta_g=0$ achieve optimal stego image quality and perfect security by generating pure Gaussian noise; however, this comes at the expense of reduced message extraction accuracy. Furthermore, GSD suffers from deteriorated visual quality and security in low-resolution (e.g., $32{\times}32$) scenarios, because the Central Limit Theorem fails in low-dimensional spaces, causing the noise $\mathbf{g}_s$ to deviate from a standard Gaussian distribution. However, as dimension increases, this problem is gradually alleviated. Despite achieving state-of-the-art extraction accuracy, GSD's constrained embedding capacity (3 bpp maximum) limits its practical use in steganography. Alternatively, MB, MC, and Multi-bits fail to maintain their claimed security when evaluated against \cite{wkk}, with detection accuracies often exceeding 95$\%$ at high payloads due to significantly compromised Gaussianity. These methods also lack flexibility for varying payloads. Conversely, PA-B2G adapts flexibly between low and high payloads. It maintains comparable image quality and security by tuning $\Delta_g$ with minimal impact on extraction accuracy. Notably, the inclusion of additional quantiles in PA-B2G mode II leads to a decrease in message extraction accuracy relative to mode I, rendering the latter more robust for real-world deployment.

\noindent\textbf{Comparison with GAN- and Flow-based Methods}.
We also compare PA-B2G with two flow-based methods, GSF \cite{weiGSF} and S2IRT \cite{zhou2022secret}, and a GAN-based method, GSN \cite{wei2022generative}, on the CelebA $64\times64$ dataset with payloads ranging from 3 to 9 bpp. GSF and S2IRT utilize the same pre-trained Glow model \footnote{https://github.com/rosinality/glow-pytorch}. For S2IRT, to maintain computational efficiency, we set the group number $K$ to 30 for a 3 bpp payload and to 40 for 6 bpp. At higher payloads, S2IRT requires a larger $K$, making it too time-consuming to generate 50,000 stego images; thus, we do not report its corresponding results in \Cref{tab:cmp_gf}. Given that PA-B2G in Mode I is superior to Mode II in extraction accuracy, we select it with $\Delta_g=0.08$ and $e=0.007$ for comparison. We employ the pre-trained diffusion model from DDIM \cite{ddim} but replace its sampling scheduler with the Heun ODE solver, as suggested in \Cref{sec:gstego}. As shown in \Cref{tab:cmp_gf}, PA-B2G significantly outperforms GAN- and flow-based methods, particularly in image quality and message extraction accuracy, primarily due to the powerful generation capabilities of diffusion models and our provable, adjustable message mapping. Regarding security, S2IRT achieves the best anti-detectability, as the extra perturbation key it introduces ensures that the generated noise strictly follows a pure Gaussian distribution. In contrast, GSF and GSN alter the noise distribution and are thus easily detected by UCNet.
\begin{figure*}
  \centering
  \begin{subfigure}[b]{\linewidth}
    \includegraphics[width=\linewidth]{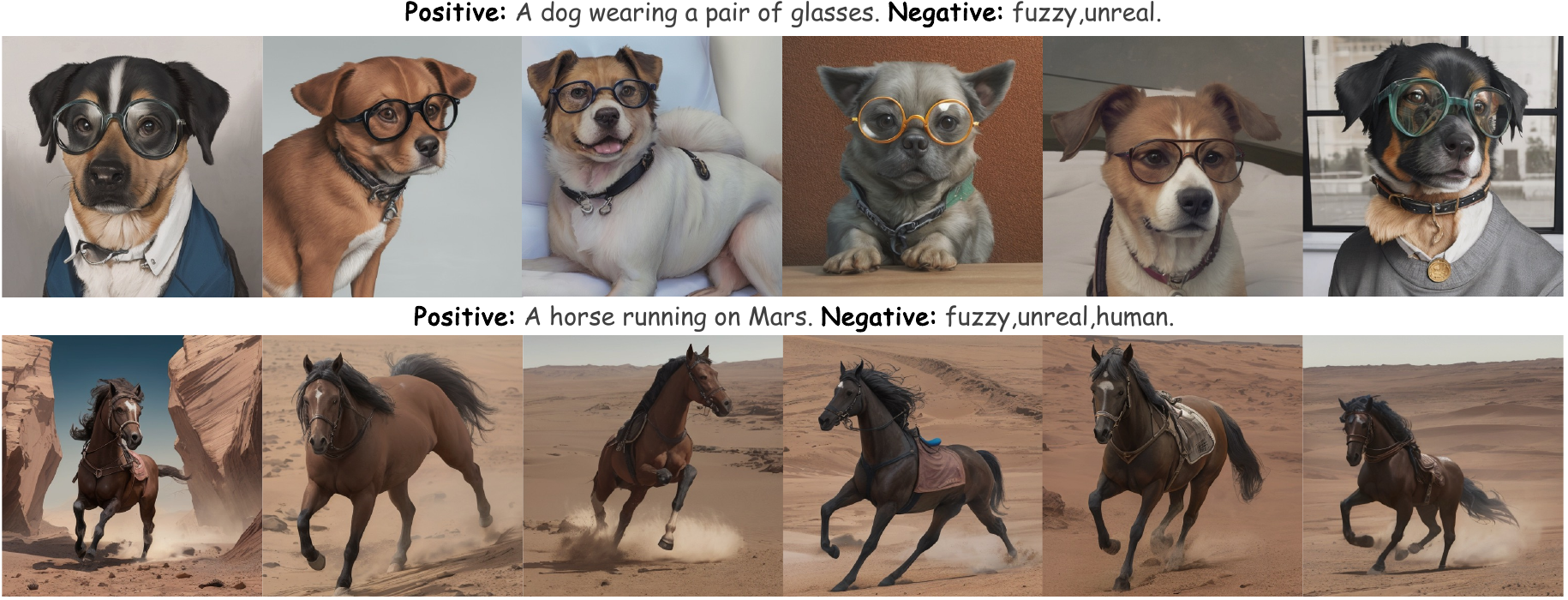}
    \caption{Text-to-image task.}
  \end{subfigure}\hfill \\
  \begin{subfigure}[b]{\linewidth}
    \includegraphics[width=\linewidth]{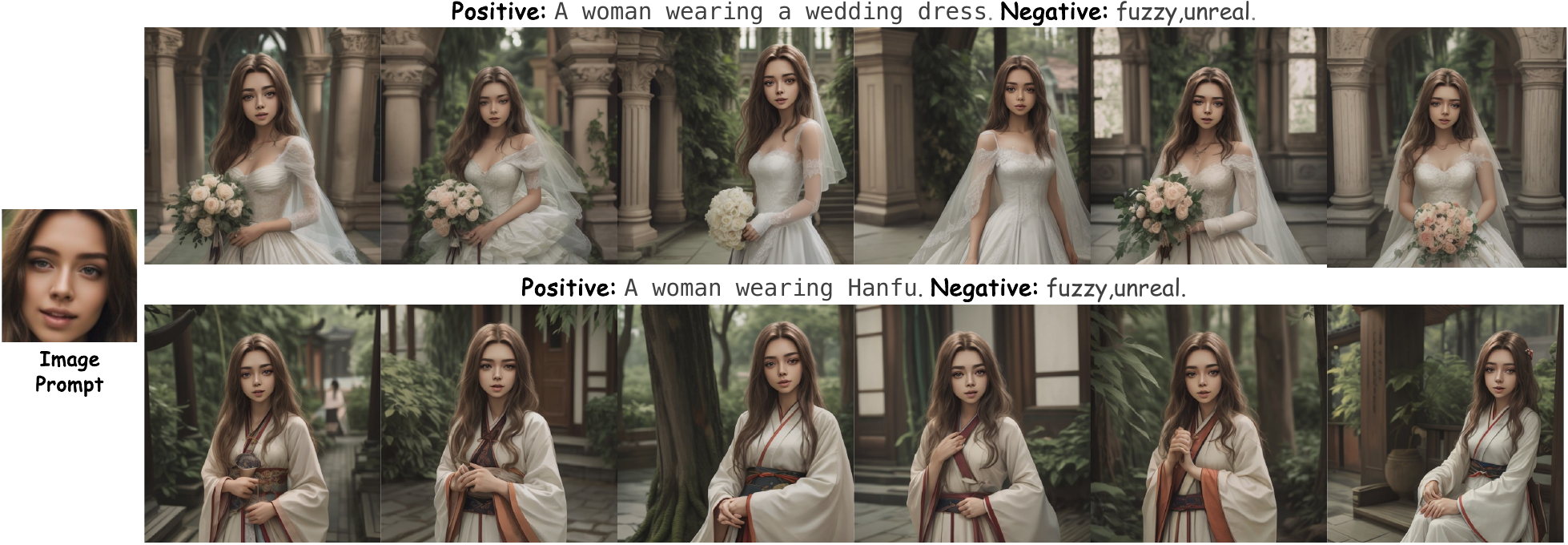}
    \caption{Image-to-image task.}
  \end{subfigure}\hfill
  \caption{Visualization results of combining PA-B2G in mode I with Stable Diffusion on text-to-image and image-to-image tasks.}
    \label{fig:stable-diffusion}
\end{figure*}

\noindent\textbf{Comparison with Prior Adjustment Strategies.}
Kim \emph{et al.} \cite{kim2025diffusion} identify a trade-off between stego image quality and extraction accuracy, proposing MN, MB, and MC as balancing options. To evaluate our adjustment strategy (\Cref{alg:pa-b2g:adjust}), \Cref{tab:cmp_mm} reports the FID score, $\overline{Acc}$, and $Acc_s$ for these methods alongside PA-B2G (modes I $\&$ II) across varying $\Delta_g$. For better intuition, \Cref{fig:trade-off} illustrates further adjustment results, confirming that image quality, security, and accuracy can be effectively balanced by tuning $\Delta_g$. This supports our analysis in \Cref{sec:exp-tradeoff} and reveals that Mode I provides a more natural, fine-grained adjustment than \cite{kim2025diffusion}. By modulating $\Delta_g$, PA-B2G can readily match the performance of MN, MB, and MC, as they represent special cases of our generalized framework.
\subsection{Diffusion Model Watermarking}\label{subsec:watermarking}
\begin{table}[!t]
\centering
\footnotesize

\setlength{\tabcolsep}{0.8mm}
\captionbox{Robustness evaluation of average message extraction accuracy (\%) under various image distortions (JPEG compression with different quality factors, random cropping with varying ratios, Gaussian noise with varying standard deviations, and Gaussian blurring with varying radii) across different payload sizes (256, 512, and 1024 bits)}{
\begin{tabular}{l ccc c ccc c ccc c ccc}
\toprule
& \multicolumn{3}{c}{JPEG} && \multicolumn{3}{c}{Random Crop} && \multicolumn{3}{c}{Gaussian Noise} && \multicolumn{3}{c}{Gaussian Blur} \\
\cmidrule{2-4} \cmidrule{6-8} \cmidrule{10-12} \cmidrule{14-16}
Payload & 90 & 75 & 50 && 0.1 & 0.3 & 0.5 && 0.01 & 0.03 & 0.05 && 2 & 3 & 4 \\
\midrule
256 bits  & 100.00\% & 99.89\% & 99.46\% && 99.98\% & 99.04\% & 87.05\% && 99.85\% & 98.03\% & 94.45\% && 99.94\% & 99.25\% & 96.64\% \\
512 bits  & 99.80\%  & 99.27\% & 98.20\% && 99.51\% & 94.20\% & 78.25\% && 98.79\% & 94.95\% & 85.58\% && 99.26\% & 96.58\%& 91.48\% \\
1024 bits & 98.60\%  & 97.30\% & 94.45\% && 97.48\% & 87.88\% & 71.22\% && 96.64\% & 88.86\% & 81.93\% && 96.04\% & 90.76\%& 83.43\% \\
\bottomrule
\end{tabular}}
\label{tab:watermark}
\end{table}
In this section, we further validate the effectiveness of PA-B2G in model I for watermarking diffusion models. We employ Dreamshaper 7, which is fine-tuned on Stable Diffusion 1.5, as the generative backbone, and utilize a first-order DDIM solver with 50 sampling steps for stego image generation. We evaluate three effective payloads—256, 512, and 1024 bits—and report their corresponding average message extraction accuracy ($\overline{Acc}$) over 1000 stego images against various lossy processing techniques in \Cref{tab:watermark}. These techniques include JPEG compression with different quality factors, random cropping with varying ratios, Gaussian noise with varying standard deviations, and Gaussian blurring with varying radii. For PA-B2G, the error tolerance $e$ is set to 0.02, code length $l$ to 1, and $\Delta_g$ to 0.05. 

As reported in \Cref{tab:watermark}, PA-B2G exhibits robust resistance to nearly all lossy processing when the effective payload is 256 bits. Notably, even with a cropping ratio of 0.5, PA-B2G still guarantees an extraction accuracy of 87.05\%. While the extraction accuracy gradually decreases as the payload increases, PA-B2G with a 1024-bit payload still effectively resists JPEG compression ($\overline{Acc} > 94\%$), which is a common operation on social media platforms. In addition to extraction accuracy, we present the visual generation results for text-to-image and image-to-image tasks in \Cref{fig:stable-diffusion}. As shown, the stego images produced by PA-B2G exhibit visual quality indistinguishable from normally generated images. These experimental results demonstrate that PA-B2G can be effectively extended to the watermarking of diffusion models.
\begin{table}[!t]
    \centering
    \captionbox{Average time required to generate noise $\mathbf{g}_s$ of different sizes using PA-B2G in modes I and II with varying $\Delta_g$.}{
	\begin{tabular}{lcc c cc c cc}
	\toprule
	 & \multicolumn{2}{c}{$32\!\times\!32\!\times 3$} && \multicolumn{2}{c}{$64\!\times\!64\!\times 3$} && \multicolumn{2}{c}{$128\!\times \!128\!\times 3$} \\
  \cmidrule{2-3}\cmidrule{5-6}\cmidrule{8-9}
      Method & 3 bpp & 6 bpp && 3 bpp & 6 bpp && 3 bpp & 6 bpp \\ \midrule
        PA-B2G$^1$(.00)       & 0.004s  & 0.005s         && 0.015s &0.016s           && 0.061s              & 0.084s         \\
        PA-B2G$^1$(.04)       & 0.018s  & 0.022s         && 0.054s &0.089s           && 0.193s              & 0.306s         \\
        PA-B2G$^1$(.08)       & 0.043s  & 0.049s         && 0.148s &0.222s           && 0.520s              & 0.871s         \\
        PA-B2G$^2$(.00)       & 0.005s  & 0.007s         && 0.025s &0.029s           && 0.084s              & 0.119s         \\
        PA-B2G$^2$(.04)       & 0.020s  & 0.025s         && 0.071s &0.074s           && 0.217s              & 0.264s         \\
        PA-B2G$^2$(.08)       & 0.046s  & 0.057s         && 0.171s &0.177s           && 0.725s              & 0.784s         \\
 \bottomrule
	\end{tabular}}
	\label{tab:tconsp}
\end{table}
\subsection{Computational Complexity}
This section evaluates the computational complexity of PA-B2G in modes I and II. The complete message-hiding pipeline encompasses four stages: mapping the secret bit sequence $\mathbf{b}$ to noise $\mathbf{g}_s$, reverse ODE solving, forward ODE solving, and recovering $\mathbf{b}$ from $\mathbf{g}_s$. \Cref{tab:tconsp} presents the average execution time for the initial mapping stage, with the error tolerance $e$ set to $0.0185$. While the computational overhead scales with noise dimensionality and $\Delta_g$, it can be mitigated by increasing the correction step $\Delta_c$ or error tolerance $e$. Notably, PA-B2G consistently completes this initial mapping in one second, regardless of the mode or noise dimensions. While the time consumption of the ODE solving processes also depends on specific diffusion architectures and sampling steps, it is considered beyond the scope of this paper. The time required for message recovery is marginal compared to the mapping process and is thus omitted.

%% file: Section/6_conclusion.tex
\section{Conclusion}
\label{sec:conclusion}
We analyze the inherent trade-off among stego-image quality, steganographic security, and message extraction accuracy in DM-GIS.
Based on this analysis, we propose PA-B2G, a provably invertible bit-to-Gaussian mapping. By integrating with off-the-shelf diffusion models, PA-B2G can establish a theoretically invertible transform between secret bitstreams and stego images. Experimental results demonstrate that our DM-GIS approach supports arbitrary-length payloads while achieving highly competitive performance in image quality, security, and extraction accuracy.
Furthermore, its robustness against lossy processing in watermarking applications validates its practical effectiveness. In future work, we plan to extend PA-B2G to broader diffusion-based generation tasks, such as video generation.